\documentclass[showpacs,twocolumn,superscriptaddress]{revtex4}

\usepackage{graphicx}
\usepackage{dcolumn}

\usepackage{amsmath}
\usepackage{amssymb}
\usepackage{bm}
\usepackage{color}
\usepackage[normalem]{ulem}
\usepackage[dvipsnames]{xcolor}
\usepackage{hyperref}
\hypersetup{
	colorlinks=true, 
	linkcolor=blue,  
	citecolor=cyan,  
}
\newcommand{\beqn}{\begin{eqnarray}}
\newcommand{\eeqn}{\end{eqnarray}}

\begin{document}
	
\title{Kerr-Taub-NUT spacetime to explain the jet power and the radiative efficiency of astrophysical black holes}

	\author{Bakhtiyor Narzilloev}
\email[]{nbakhtiyor18@fudan.edu.cn}
\affiliation{Center for Field Theory and Particle Physics and Department of Physics, Fudan University, 200438 Shanghai, China }
	\affiliation{School of Engineering, Central Asian University, Tashkent 111221, Uzbekistan}
	\affiliation{Ulugh Beg Astronomical Institute, Astronomy St.  33, Tashkent 100052, Uzbekistan}
\affiliation{Tashkent Institute of Irrigation and Agricultural Mechanization Engineers, Kori Niyoziy, 39, Tashkent 100000, Uzbekistan}

\author{Ahmadjon~Abdujabbarov}
	\email[]{ahmadjon@astrin.uz}
	\affiliation{Ulugh Beg Astronomical Institute, Astronomy St.  33, Tashkent 100052, Uzbekistan}
		\affiliation{Tashkent State Technical University, Tashkent 100095, Uzbekistan}
	\affiliation{Institute of Nuclear Physics, Ulugbek 1, Tashkent 100214, Uzbekistan}
	\affiliation{Institute of Theoretical Physics, National University of Uzbekistan, Tashkent 100174, Uzbekistan}

\author{Bobomurat Ahmedov}
\email[]{ahmedov@astrin.uz}
\affiliation{Ulugh Beg Astronomical Institute, Astronomy St. 33, Tashkent 100052, Uzbekistan}
\affiliation{Institute of Theoretical Physics, National University of Uzbekistan, Tashkent 100174, Uzbekistan}
\affiliation{Tashkent Institute of Irrigation and Agricultural Mechanization Engineers, Kori Niyoziy, 39, Tashkent 100000, Uzbekistan}
\author{Cosimo Bambi}
\email[]{bambi@fudan.edu.cn}
\affiliation{Center for Field Theory and Particle Physics and Department of Physics, Fudan University, 200438 Shanghai, China }
	\date{\today}

\begin{abstract}
In this work, we investigate the electromagnetic energy released by astrophysical black holes within the Kerr-Taub-NUT solution, which describes rotating black holes with a nonvanishing gravitomagnetic charge. In our study, we consider the black holes in the X-ray binary systems GRS~1915+105, GRO~J1655-40, XTE~J1550-564, A0620-00, H1743-322, and GRS~1124-683. We show that the Kerr-Taub-NUT spacetime can explain the radiative efficiency of these sources inferred from the continuum fitting method (CFM). We also show that, in the framework of the Blandford-Znajeck mechanism, it is possible to reproduce the observed jet power. We unify the results of the two analyses for the selected objects to get more stringent constraints on the spacetime parameters. We show that, as in the case of the Kerr spacetime, the Kerr-Taub-NUT solution cannot simultaneously explain the observed jet power and radiative efficiency of GRS~1915+105.
\end{abstract}

\maketitle

\section{Introduction}

The vacuum solution of the Einstein field equation describing a rotating black hole with a nonzero gravitomagnetic charge is known as the Kerr-Taub-NUT spacetime~\cite{Miller73}. Being a stationary and axisymmetric solution of the field equations of general relativity, the Kerr-Taub-NUT spacetime is a special case of the Plebanski and Demianski solution~\cite{ref25}, but its uniqueness is that it admits separable Hamilton-Jacobi and Klein-Gordon equations~\cite{Dadhich:2001sz}. On the other hand, the Kerr-Taub-NUT solution can be seen as a generalization of the Kerr spacetime, which describes a rotating black hole with vanishing gravitomagnetic charge. It has been pointed out that recent observations of the X-ray binary GRO~J1655-40 may be interpreted with the presence of a nonvanishing gravitomagnetic monopole moment in its black hole  \cite{ref0}. From such a suggestive possibility, here we explore the astrophysical processes of radiative efficiency of accreting  matter onto a black hole and of relativistic jets with the aim to get some constraints on the gravitomagnetic charge of specific objects. We note that the properties of various black hole solutions have been extensively studied in our previous works \cite{Hakimov17,   Narzilloev22a, Narzilloev22b, Narzilloev22c, Narzilloev23, Narzilloev23a, Narzilloev2023b, Narzilloev2023c, Narzilloev2023d, Narzilloev2023e}. The Blandford-Znajeck mechanism of energy extraction in a general axially-symmetric black-hole spacetime has been recently studied in \cite{Konoplya2021qll}.


Newmann, Unti and Tamburino \cite{r4} were the first to propose a stationary and spherically symmetric \cite{r5, r6} vacuum solution of the Einstein field equations including an additional parameter responsible for the gravitomagnetic monopole charge, also called the NUT parameter. Demianski and Newman showed that the NUT spacetime is produced by a  so called ``dual mass'' \cite{r9} and can be interpreted as a gravitomagnetic charge.  In order to understand the nature of the gravitomagnetic monopole, one can consider the analogy with Dirac's magnetic monopole~\cite{r10, r11}. In particular, the author of Ref.~\cite{r12} suggested to interpret the NUT parameter as a linear source of pure angular momentum \cite{r7, r13} that can be understood as a massless rotating rod. The gravitomagnetic charge should be a conserved quantity in common astrophysical processes, so it is conserved in the merger of two black holes or in the accretion process of matter onto a black hole. Since ordinary matter from nearby stars or the interstellar medium has vanishing gravitomagnetic charge, the accretion process should reduce the gravitomagnetic charge to mass ratio of a black hole with an initially non-vanishing gravitomagnetic charge. This may significantly limit the gravitomagnetic charge of supermassive black holes in galactic nuclei, as it is thought that their mass is mainly the result of the accretion process over billions of years, while the impact of the accretion process is thought to be negligible for stellar-mass black holes. The authors of Ref.~\cite{r6} suggested that the signature of a gravitomagnetic monopole may be found in the spectra of supernovae, quasars, and active galactic nuclei \cite{r6, r14}. The effects of a gravitomagnetic monopole momentum on light rays was studied in Refs.~\cite{r15, r16}. The energy of plasma magnetosphere of neutron stars would strongly depend on the NUT parameter, as shown in \cite{r17}.

In this paper, we investigate the radiative efficiency and the power of relativistic jets in the background spacetime of Kerr-Taub-NUT black holes. Relativistic jets are commonly observed in active galactic nuclei and black hole X-ray binaries (microquasars) and are thought to start near the black hole event horizon~\cite{ref77}. In such a case, the spacetime geometry around the black hole can have a strong impact on the power of relativistic jets.  At the same time, a possible nonvanishing gravitomagnetic charge can alter the position of the innermost stable circular orbit (ISCO) and the thermal spectrum of the possible accretion disk. In this paper, we will use the Novikov-Thorne model to interpret astrophysical data~\cite{ref70,ref78}. It is worth mentioning here that the idea to combine the results of the two observations to investigate the spacetime geometry around black hole candidates is not novel and has been developed by several authors. For example, the relation between jet power and spin was originally found in \cite{ref84} and \cite{ref91}. The combination of this finding with the CFM spin measurements to test the spacetime geometry was discussed in  \cite{ref78} and \cite{PhysRevD86123013}. 

The present paper has the following structure. In Section~\ref{section4.2}, we briefly review the Kerr-Taub-NUT metric and its main properties. In Section~\ref{section4.3}, we give theoretical aspects of observables that can be used to get constraints on the spacetime parameters. In Section~\ref{section4.4}, we get constraints on the spacetime parameters of the Kerr-Taub-NUT metric for selected black hole candidates. We summarize our main results in Section~\ref{section4.6}. Throughout the paper we use natural units in which $G=c=1$.

\section{Kerr-Taub-NUT spacetime \label{section4.2}}

The line element of the Kerr-Taub-NUT solution describing a black hole of mass $M$, gravitomagnetic charge $l_*$, and rotational parameter $a_*$ reads~\cite{Abdujabbarov08}
\begin{eqnarray}
\nonumber
ds^2=&-&\frac{1}{\Sigma}\left(\Delta-a_*^2\sin^2\theta \right)dt^2+\Sigma\left( \frac{1}{\Delta}dr^2+d\theta^2\right)\\\nonumber
&+&\frac{1}{\Sigma}\left[ (\Sigma+a_*\chi)^2\sin^2\theta-\chi^2\Delta \right]d\phi^2 \\
&+&\frac{2}{\Sigma}\left( \Delta \chi -a_*(\Sigma+a_*\chi)\sin^2\theta\right)d\phi dt \ ,
\end{eqnarray}
where $\Delta$, $\Sigma$, and $\chi$ are defined as 
\begin{eqnarray}
\Delta&=&r^2+a_*^2-l_*^2-2Mr \ ,\\
\Sigma&=&r^2+(l_*+a_*\cos\theta)^2 \ ,\\
\chi&=&a_*\sin^2\theta-2l_*\cos\theta \ .
\end{eqnarray}
One can determine the radial coordinate of the event horizon from the condition $\Delta=0$, which has the exact analytical solution
\begin{equation}\label{eh}
    r_H=M+\sqrt{M^2-a_*^2+l_*^2} .
\end{equation}
The outer radius of the ergoregion is given by
\begin{eqnarray}
r_e=M+\sqrt{M^2-a_*^2 \cos^2\theta+l_*^2}.
\end{eqnarray}
As we can see from the equations above, on the poles ($\theta=0$ or $\pi$) the radius of the ergoregion reaches the event horizon. However, on the equatorial plane ($\theta=\pi/2$) the ergoregion radius depends on the NUT parameter and the mass, $r_e|_{\theta=\frac{\pi}{2}}=M+\sqrt{M^2+l_*^2}$, while in the Kerr spacetime it is $r_e|_{\theta=\frac{\pi}{2}}=2 M$. Setting the denominator of the metric component $g_{tt}$ to zero, one finds 
$$r=0\,\,\, \text{and}\,\,\, \theta=\cos^{-1}(-l_*/a_*)\, .$$ 
This is the location of the singularity in the Kerr-Taub-NUT spacetime. Note that the spacetime becomes singularity free when $l_*>a_*$, namely we have a regular black hole when $l_*>a_*$. The Kerr-Taub-NUT spacetime describes a black hole when $|a_*| \le \sqrt{M^2+l_*^2}$ and a naked singularity when $a_* > \sqrt{M^2+l_*^2}$. In what follows, we will often use the dimensionless spin parameter $a=a_*/M$ and the dimensionless NUT parameter $l=l_*/M$ for simplicity (note that in the literature the notation is normally different, where $a_*$ is dimensionless and $a = a_* M$). 
In principle, for a Kerr-Taub-NUT black hole, the values of $a$ and $l$ can be arbitrarily large (but, of course, they must satisfy the condition $|a| \le \sqrt{1+l^2}$). However, in this work we are going to take the values of the gravitomagnetic charge to be $l\le1$ i.e. we assume that selected black hole candidates do not have very big gravitomagnetic charge (later we will see that for the source GRO J1655-40 it fails to explain the jet power of the source for even smaller values of $l$) which, in turn, restricts the spin of a black hole to be in the range $|a|\le\sqrt{2}$.

\section{theoretical aspects \label{section4.3}}

\subsection{Radiative efficiency of the system}

In this subsection, we discuss the continuum spectrum emitted by a black hole accretion disc and we consider the Novikov-Thorne model~\cite{ref70}. The model assumes that the disc is geometrically thin, so particles move on or very close to the equatorial plane. Such particles have almost circular trajectories. The tiny radial motion caused by viscous forces makes particles to move along spiral-like trajectories and eventually they fall onto the black hole. 

Since the gravitational force is supposed to dominate the gas motion over the gas pressure, we can assume that the particles of the disc follow circular geodesic orbits. As the gas falls onto the gravitational well of the central massive object, it loses energy and angular momentum. A part of this energy is converted into electromagnetic radiation

The Novikov-Thorne accretion disc is geometrically thin and optically thick and there is no trapped heat. The gas is in local thermal equilibrium and every point on the disc has a blackbody spectrum. The whole disc has a multi-temperature blackbody spectrum and the emission is normally peaked in the soft X-ray band for stellar-mass black holes and in the UV band for supermassive black holes (see, e.g.~\cite{ref70,ref79,Bambi17e, 2021SSRv..217...65B}).

The thermal spectrum of the accretion disc is very sensitive to the location of the inner edge of the disc. If we assume that the inner edge is at the ISCO radius and we have independent estimates of the black hole mass, distance, and the inclination angle of the disc, we can fit the data and infer the location of the ISCO radius~\cite{Zhang_1997}. Note that the ISCO radius depends on the specific background metric. The ISCO radius can be inferred from the effective potential of a particle orbiting a black hole as follows. Assuming an axially symmetric and stationary spacetime with the metric $g_{\mu \nu}$ written in the canonical form, from the normalization of the particle 4-velocity $u_\mu u^\mu = -1$ we have
\begin{eqnarray}
g_{rr} u_r^2 + g_{\theta\theta} u_\theta^2 = V_{\rm eff} ,
\end{eqnarray}
where the effective potential $V_{\rm eff}$ is~\cite{Bambi17e}
\begin{eqnarray}
V_{\rm eff}=\frac{E^2 g_{\phi \phi}+2 E L g_{t \phi}+L^2 g_{tt}}{g_{t\phi}^2-g_{tt}g_{\phi \phi}}-1,
\end{eqnarray}
and where $E$ and $L$ are, respectively, the specific energy and the specific angular momentum of the orbiting massive particle. In terms of the metric components, these quantities take the following form
\begin{eqnarray}
E=\frac{-g_{tt}-\Omega g_{t\phi}}{\sqrt{-g_{tt}-2\Omega g_{t\phi}-\Omega^2 g_{\phi \phi}}}\ ,
\end{eqnarray}
for the energy, and
\begin{eqnarray}
L=\frac{\Omega g_{\phi \phi}+g_{t \phi}}{\sqrt{-g_{tt}-2\Omega g_{t\phi}-\Omega^2 g_{\phi \phi}}}\ ,
\end{eqnarray}
for the angular momentum. Here $\Omega = d\phi/dt$ is the angular velocity of the particle~\cite{Bambi17e}
\beqn
\Omega=\frac{d\phi}{dt} = \frac{-g_{t\phi,r}\pm\sqrt{\{-g_{t\phi,r}\}^2-\{g_{\phi \phi, r}\} \{g_{tt,r}\}}}{g_{\phi \phi, r}}\ 
\eeqn
and $g_{\mu\nu,\rho} \equiv \partial_{\rho}g_{\mu\nu}$. To calculate the ISCO radius, one needs to solve the following set of equations
\begin{eqnarray}\label{isco_eqn}
V_{\rm eff}(r)=0\ ,\quad
V_{\rm eff}'(r)=0\ ,\quad
V_{\rm eff}''(r)=0\ ,
\end{eqnarray}
where the apostrophe $'$ denotes a derivative with respect to $r$. We see here that such conditions involve the spacetime metric components and this will allow us to constrain the parameters of the spacetime around the central compact object by measuring the ISCO from the continuum spectrum. For example, if the spacetime metric is described by the Kerr solution, we can estimate the black hole spin~\cite{Zhang_1997}. This is the so-called Continuum Fitting Method (CFM), which has been extensively used in the past two decades to estimate the spin of stellar-mass black holes~\cite{McClintock:2013vwa}. 

\begin{figure*}[t!]
	\begin{center}
		\includegraphics[width=0.49\linewidth]{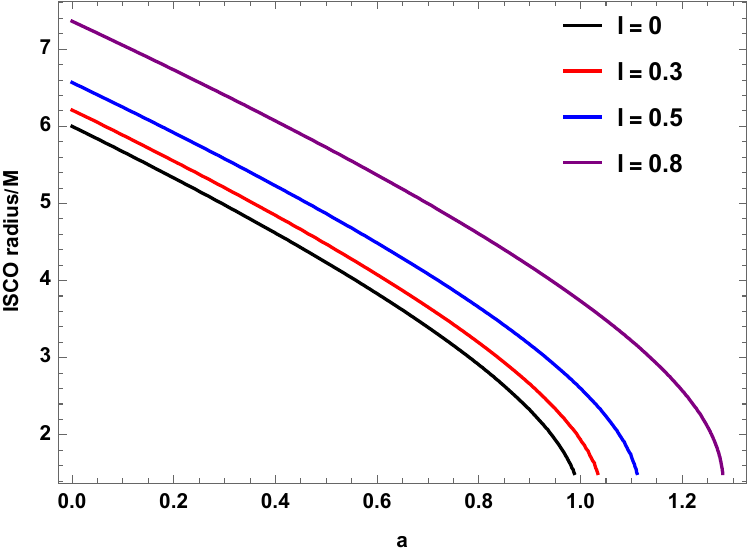}
		\includegraphics[width=0.49\linewidth]{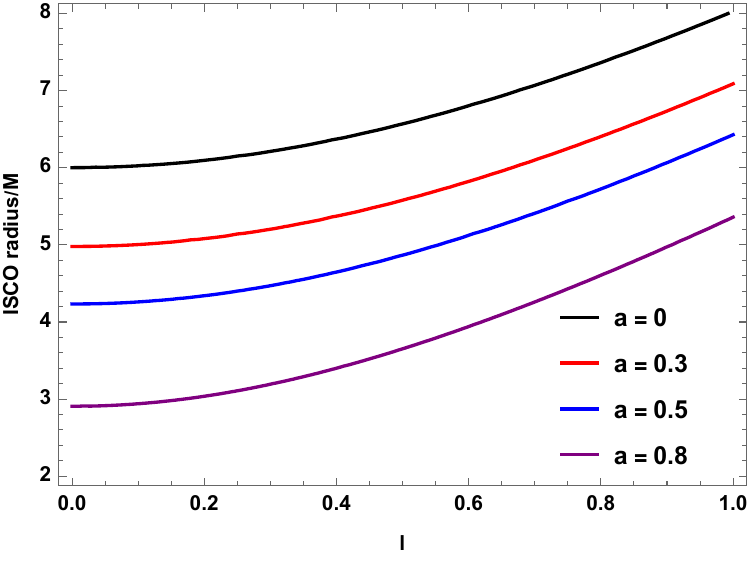}
	\end{center}
	\caption{Effect of the dimensionless spin $a$ and the dimensionless gravitomagnetic charge $l$ in the Kerr-taub-NUT spacetime on the ISCO radius of test particles. \label{isco}}
\end{figure*}

The dependence of the ISCO radius from the parameters $a$ and $l$ of the Kerr-taub-NUT spacetime is shown in Fig. \ref{isco}. If the spin parameter increases/decreases, the ISCO radius decreases/increases. The opposite effect is produced by the gravitomagnetic charge: if the gravitomagnetic charge increases/decreases, the ISCO radius also increases/decreases.

\begin{figure}[b]
	\begin{center}
		\includegraphics[width=0.95\linewidth]{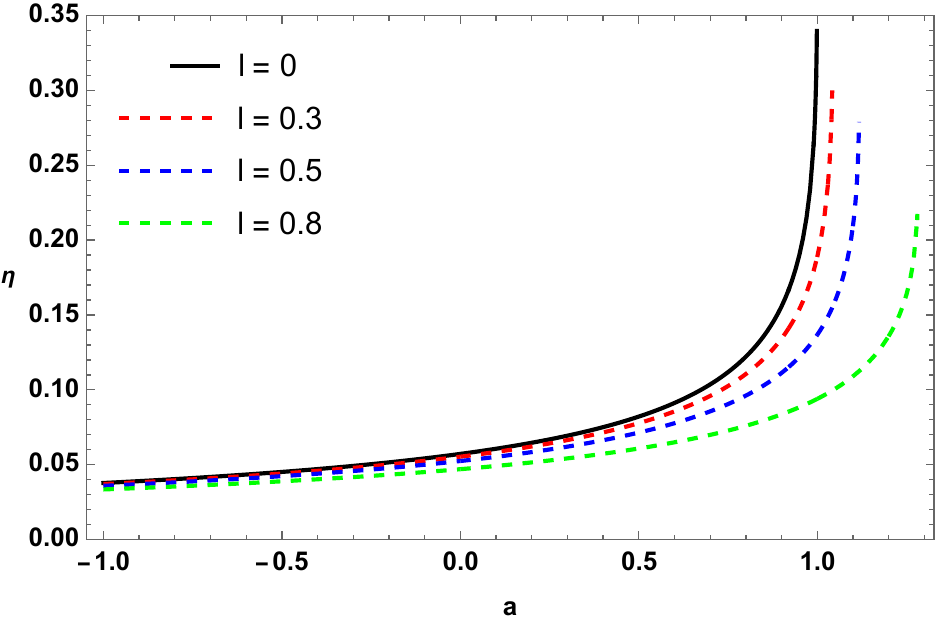}
	\end{center}
	\caption{Radiative efficiency $\eta$ of Kerr-Taub-NUT black holes as a function of the spin parameter $a$ for the different values of the NUT parameter $l$. \label{eff}}
\end{figure}

The radiative efficiency of a Novikov-Thorne accretion disc, $\eta$, is equal to the binding energy of a particle orbiting the black hole at the ISCO radius, so 
\begin{eqnarray}
\eta=1-E_{isco}\, , \label{etais}
\end{eqnarray}
where $E_{isco}$ is the specific energy of the particle at the ISCO radius. $\eta$ thus depends on the spacetime metric. In the Kerr spacetime, it is a function of the rotational parameter $a$ only. In the Kerr-Taub-NUT spacetime, $\eta$ is determined by the values of $a$ and $l$. Fig.~\ref{eff} shows how the radiative efficiency $\eta$ changes for different values of the black hole spin parameter $a$ and of the NUT parameter $l$. Increasing the NUT parameter $l$, we decrease the radiative efficiency $\eta$ and the effect is larger for faster rotation of black holes. At first approximation, black holes with a Novikov-Thorne accretion disc with the same radiative efficiency have the same thermal spectrum \cite{Kong14}, and such a result can be used to estimate the parameters of the spacetime metric (this point will be clarified later).

\subsection{Relativistic jets }

\textcolor{black}{
One may distinguish two types of jets from microquasars~\cite{ref68}:}
\begin{itemize}
    \item Steady, non-relativistic jets.  They are common during the hard-state~\cite{ref83} and are observed over a large interval of accretion luminosities.
    
    \item Transient or ballistic jets occurring when the luminosity of the source is around its Eddington limit during the transition from the hard to soft state. Usually this type of jets has a relativistic nature and is believed to be launched from near the horizon~\cite{ref77}. 

\end{itemize}

\begin{figure}
	\begin{center}
		\includegraphics[width= 0.9\linewidth]{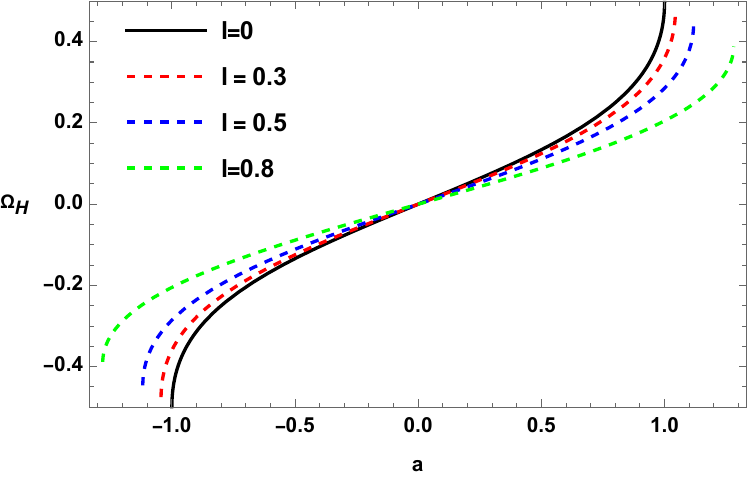}
	\end{center}
	\caption{\textcolor{black}{Angular velocity at the event horizon as a function of the spin parameter $a$ for a few different values of the NUT parameter $l$. \label{W_a}}}
\end{figure}

Here we consider the second type of jets in order to extract  information on black hole spin and gravitomagnetic charge. Despite the large number of attempts to describe the mechanism of generation of relativistic transient jets~\cite{ref85, ref87}, there is still no consistent model to explain observations. Here we employ the mechanism of energy extraction from a black hole proposed by Blandford and Znajeck, which can be used for any axial-symmetric spacetime metric. This model considers the formation of relativistic jets powered by the rotational energy of the black hole through the magnetic field of the current-carrying accretion disc. The total energy-momentum tensor contains the electromagnetic field only and other contributions are neglected: 
\beqn
T_{\mu \nu}^{tot} \simeq T_{\mu \nu}^{EM}=F_{\mu \alpha} F^\alpha_\nu-\frac{1}{4}g_{\mu \nu} F_{\alpha \beta} F^{\alpha \beta}\ .
\eeqn
In such a case, the conservation equation reduces to
\beqn
\nabla^\mu T_{\mu \nu}^{EM}=0\ ,
\eeqn
where $F_{\mu\nu} = A_{\nu,\mu} - A_{\mu,\nu}$ is the electromagnetic field tensor corresponding to the four potential $A_\mu$. For a force-free magnetosphere, one can easily write the following expression
\beqn
\frac{A_{t,r}}{A_\phi,r}=\frac{A_{t,\theta}}{A_{\phi,\theta}}=-\omega(r,\theta)\ ,\label{ffree}
\eeqn
where $\omega(r,\theta)$ can be interpreted as an electromagnetic angular velocity~\cite{ref69}. Using the condition~(\ref{ffree}) for the axisymmetric and time-independent four potential of the electromagnetic field, one can express $F_{\mu\nu}$ in the following form
\beqn
F_{\mu \nu}=\sqrt{-g} \left(
\begin{array}{cccc}
 0 & - \omega B^{\theta }   &\omega B^r   & 0 \\
 \omega B^{\theta }  & 0 & B^{\phi } & -B^{\theta } \\
 -\omega B^r  & -B^{\phi } & 0 & B^r \\
 0 & B^{\theta } & -B^r & 0 \\
\end{array}
\right)\ .
\eeqn
The power of the relativistic jets within this model has the following form~\cite{ref69}: %
\beqn
P_{BZ}=4 \pi \int_0^{\pi/2} \sqrt{-g} T_t^r d\theta \ , 
\eeqn
where $T_t^r$ is the radial component of the Poynting flux and is assumed that the jet is launched at the event horizon. The radial component of the Poynting flux is given by 
\beqn
T_t^r=2 r_HM \sin^2\theta (B^r)^2 \omega [\Omega_H-\omega]|_{r=r_H}\, ,
\eeqn
with the angular velocity $\Omega_H$ evaluated at the event horizon $r_H$ and given by the expression $$\Omega_H=-\frac{g_{t\phi}}{g_{\phi \phi}}|_{r_H}=\frac{2 a_* \left(l_*^2+M r_H\right)}{a_*^2 \left(3 l_*^2+r_H (2 M+r_H)\right)+\left(l_*^2+r_H^2\right)^2}\, .$$ 
It is worth noting here that the original work by Blandford and Znajek discussed this phenomenon in the slow-rotation limit, valid for $a$ close to zero, finding that the jet power had to be proportional to $a^2$~\cite{ref69}. In Ref.~\cite{ref89}, Tchekhovskoy et al. extended the original result to almost the entire range of spin parameter of Kerr BH, finding that the jet power in the Blandford-Znajeck model should be proportional to the square of $\Omega_H$ at the leading order \cite{ref89,Camilloni:2022kmx}
\beqn
P_{BZ}=k \Phi_{tot}^2 \Omega_H^2\ , \label{PBZ}
\eeqn
where $k = 1/6\pi$ for a split monopole field profile and $k = 0.044$ for a paraboloidal one \cite{ref69}. The results in \cite{ref89,Camilloni:2022kmx} are obtained in the Kerr metric, but studies of the Blandford-Znajeck mechanism in other theories of gravity showed that only higher order corrections depend on the specific gravity model and the formula above at the leading order does not change~\cite{Camilloni:2023wyn}. In Eq.~(\ref{PBZ}), $\Phi_{tot}$ is the magnetic flux and is given by 
\beqn
\Phi_{tot}=2 \pi \int_0^{\pi} \sqrt{-g} |B^r| d\theta\ .
\eeqn
In Fig.~\ref{W_a} the dependence of the angular velocity $\Omega_H$ from the spin $a$ for various values of the gravitomagnetic charge is presented. The change of angular velocity $\Omega_H$, due to the presence of the NUT parameter has an impact on the power of relativistic jets.

\section{Constraints from observational data \label{section4.4}}

Since the radiative efficiency (\ref{etais}) is sensitive to the spacetime metric, its measurement can be used to estimate/constrain the black hole parameters of the corresponding theory of gravity. In what follows, we will consider the following objects that will be interpreted as Kerr-Taub-NUT black holes: GRS1915+105, GROJ1655-40, XTEJ1550-564, A0620-00, H1743-322, and GRS1124-683 ~\cite{ref84,ref90,ref91}. Table~\ref{Table1} shows the measurements of some properties of these systems reported in the literature. The estimates of the black hole spin $a$ (and the derived estimate of the Novikov-Thorne radiative efficiency $\eta$) are all obtained assuming the Kerr metric.

\begin{widetext}
\begin{center}
\begin{table}[!ht]
\centering 
    \caption{Parameters of the black hole binaries analyzed in this work. The radiative efficiency $\eta$ is obtained from the spin measurement by using Eq.~(\ref{etais}) for the Kerr metric.}\label{Table1}
    \begin{tabular}{|c|c|c|c|c|c|}
    \hline
    BH Source & $M \, (M_\odot)$ & $D \, (kpc)$ & $i^o$ & $a$ & $\eta$ 
    \\ 
    \hline
  A0620-00 & $6.61 { \pm 0.25}$ & $1.06 { \pm 0.12}$ & $51.0 { \pm 0.9}$ & $0.12 { \pm 0.19}$~\cite{ref98} & $0.061^{+0.009}_{-0.007}$ 
  \\
    \hline
    H1743-322 & $8.0$ & $8.5 {  \pm 0.8}$ & $75.0 { \pm 3.0}$ & $0.2 { \pm 0.3}$ \cite{ref100}& $0.065^{+0.017}_{-0.011}$ 
    \\ 
    \hline
   XTEJ1550-564 & $9.10 { \pm 0.61}$ & $4.38 { \pm 0.5}$ & $74.7 { \pm 3.8}$ & $0.34 { \pm 0.24}$ \cite{steiner2011spin} & $0.072^{+0.017}_{-0.011}$ 
   \\ 
    \hline
   GRS1124-683 & $11.0^{+2.1}_{-1.4}$ & $4.95^{+0.69}_{-0.65}$ & $43.2^{+2.1}_{-2.7}$ & $0.63^{+0.16}_{-0.19}$ \cite{ref109} & $0.095^{+0.025}_{-0.017}$ 
   \\ 
    \hline
   GROJ1655-40 & $6.30 { \pm 0.27}$ & $3.2 { \pm 0.5}$ & $70.2 { \pm 1.9}$ & $0.7 { \pm 0.1}$ \cite{Shafee06}& $0.104^{+0.018}_{-0.013}$ 
   \\ 
    \hline
  GRS1915+105 & $12.4^{+1.7}_{-1.9}$ & $8.6^{+2.0}_{-1.6}$ & $60.0 { \pm 5.0}$ & $a_{*} > 0.98$ \cite{ref118} & $\eta { > 0.234}$ 
  \\ 
    \hline
    \end{tabular}
  \end{table}
  \end{center}
\end{widetext}

Here we use the procedure described in~\cite{ref84,ref91} to evaluate the jet power of the six objects of our study. We can think of a bipolar radio jet as a symmetrical pair of plasmoids. These plasmoids emit radiation in an isotropic manner and have a thin optical structure. They expand outward from the core source at a relativistic bulk velocity $\beta$. The ratio between the observed and emitted flux density for each individual jet can be written as $$S_\nu/S_{\nu,0}=\delta^{3-\alpha} .$$ In this context, the Doppler factor is denoted as $\delta$, and the radio spectral index is represented by $\alpha$. The Doppler factor for the brighter jet, which is approaching, can be straightforwardly written in terms of $\beta$, the Lorentz factor $\Gamma$, and the inclination angle $i$ of the jet as follows: $$\delta=(\Gamma [1-\beta \cos i])^{-1} .$$ In the case of the main emission source, specifically the approaching jet, the observed intensity surpasses the emitted intensity at lower inclinations, while the opposite is true for higher inclinations. For microquasars with mildly relativistic jets, the Doppler boost becomes less than one within the intermediate range of inclinations, approximately between 35 and 55 degrees. We assume that the entire power in the transient jet is proportional to the peak at 5~GHz of the radio flux density (see Table~\ref{Table2}). In natural units, the luminosity can be written as~\cite{ref84,ref91} 
\begin{eqnarray}
P_{jet}=\left(\frac{\nu}{5~{\rm GHz}}\right) \left(\frac{S_{\nu,0}^{tot}}{\rm Jy}\right) \left(\frac{D}{\rm kpc}\right)^2 \left(\frac{M}{M_{\odot}}\right)^{-1}\, ,
\end{eqnarray}
where $S_{\nu,0}^{tot}$ is the beaming corresponding to the approaching and receding jets~\cite{ref77,ref91}. The Lorenz factor $\Gamma$ associated to the jet can be expected to be in the interval $2\leq\Gamma\leq5$. The Doppler corrected jet powers corresponding to Lorenz factor $\Gamma=2$ and $\Gamma=5$ for every source are given in Table~\ref{Table2}~\cite{ref90,ref94}. 

\begin{table}[!ht]
    \large        
    \caption{Proxy jet power values in units of kpc$^2$~GHz~Jy~$M_{\odot}^{-1}$}\label{Table2}
    \centering    
    \begin{tabular}{|c|c|c|c|}
    \hline
    BH Source & $(S_{\nu,0})_{max}^{5 GHz}$ (Jy) & $P_{jet}|_{\Gamma=2}$ & $P_{jet}|_{\Gamma=5}$\\     
    \hline
    A0620-00 & 0.203 & 0.13 & 1.6\\
    \hline
    H1743-322 &0.0346 & 7.0 & 140\\
    \hline
     XTEJ1550-564 & 0.265 & 11 & 180\\
    \hline
     GRS1124-683 & 0.45 & 3.9 & 380\\
    \hline
     GROJ1655-40 & 2.42 & 70 & 1600\\
    \hline
     GRS1915+105 & 0.912 & 42 & 660\\
    \hline
    \end{tabular}
  \end{table}

The results of Table~\ref{Table2} can be compared with the theoretical predictions, which depend on the spacetime metric. From Eq.\eqref{PBZ}, the power of the jet can be expressed as 
\beqn\label{P}
\log P=\log K+2 \log \Omega_H\, ,
\eeqn
where $K = k \Phi_{tot}^2$. 
Here, the value of $K$ can be found  by fitting the observed jet power and $\Omega_H$~\cite{ref84,ref94}. The spacetime metric enters the calculations of the jet power through the angular velocity of the event horizon $\Omega_H^2$. 
In Ref.~\cite{ref94}, the authors inferred the best-fitting values of the parameter $K$. They found $\log K = 2.94 \pm 0.22$ for the Lorentz factor $\Gamma=2$ and $\log K = 4.19 \pm 0.22$ for $\Gamma=5$ (90\% confidence level). It is worth noting that, in general, $K$ should not be a constant for every source. However, it is believed that magnetic field strength depends on mass accretion rate $\Dot{M}$ \cite{ref84, Tchekhovskoyetal.2011}. Transient jets show up during the transition from the hard to soft state and therefore the Eddington scaled mass accretion rate is similar for all the sources. Even the mass is very similar, of order of 10 Solar masses. For this reason one can take this quantity as constant for the  six BH candidates in our list. Since we take $K$ to be independent of the spacetime geometry, hereafter we use these values of $K$ to constrain the spin parameter and the NUT parameter of the Kerr-Taub-NUT spacetime from the observed jet power of the sources in Tab.~\ref{Table2}. In other words, we use Eq.~\eqref{P} to get the constraints between the black hole spins and their gravitomagnetic charges, taking into account that they enter the calculations through $\Omega_H$.

\subsection{Results}

Below we present our results for every source. 

\begin{itemize}

 \item {\bf Source A0620-00}. 
From the CFM and assuming the Kerr metric, the spin parameter of the source has been estimated $a=0.12\pm0.19$ at 68\%  confidence level (CL)~\cite{ref98}. Such a spin measurement can be rewritten as a measurement of the radiative efficiency, $\eta = 0.061_{-0.007}^{+0.009}$, which is reported in the last column in Tab.~\ref{Table1}.
As discussed in Ref.~\cite{Kong14}, at first approximation the CFM measures the radiative efficiency of the Novikov-Thorne disc of the source, and such a value can be easily translated into a constraint on the spacetime parameters for a putative non-Kerr black hole. This is how we constrain the parameters of the Kerr-Taub-NUT solution here. Fig.~\ref{1Je} shows (blue regions) the constraints on the spin parameter $a$ and the NUT parameter $l$ imposing the requirement that the Novikov-Thorne radiative efficiency is $\eta = 0.061_{-0.007}^{+0.009}$. For $l=0$, we recover $a=0.12\pm0.19$, but much higher values of the spin parameter are allowed for $l > 0$. 
The radiative efficiency is degenerate with respect to $a$ and $l$ and it is impossible to constrain the two parameters without another measurement.
\textcolor{black}{The left panel of Fig.~\ref{1Je} is for the values of  $\Gamma=2$ and right panel is for $\Gamma=5$. The solid red line in the panels corresponds to the central value of $P_{jet}$ and the dashed red lines describe the values of spacetime parameters corresponding to the $P_{jet}$ with the error of $0.3 \, dex$ around the central jet power presented in Tab.~\ref{Table2}. The shaded red regions correspond to the values of spacetime parameters within the error bars.
From panel (a) and panel (b), one can see that Kerr-Taub-NUT spacetime is efficient to explain the jet power of the source. It is shown for the case  $\Gamma=2$ that the central value of the jet power explained by the Kerr spacetime with the spin parameter $a\simeq0.05$ can also be well explained by the Kerr-Taub-NUT spacetime with the values of the parameters corresponding to the points on the red line. In the right panel, which is for the Lorentz factor of $\Gamma=5$, we see that the difference compared to the case of $\Gamma=2$ exists but it is not negligible. One can check that the increase of the gravitomagnetic charge causes to rise slightly the value of the spin parameter of a black hole  described by the Kerr-Taub-NUT spacetime to match the values of the observed jet power.}
\textcolor{black}{In Fig.~\ref{1Je} we also see that the cases when $\Gamma=2$ and $\Gamma=5$ have very similar behaviors and from the intersection of the blue shaded and red shaded regions one can state that, theoretically, such regions give the values of the spacetime parameters that can explain both observations with the Kerr-Taub-NUT spacetime within the error bars. So, from the figure we see that the radiative efficiency and the jet power of the source A0620-00 can be simultaneously well explained by Kerr-Taub-NUT spacetime with parameters in the approximate range $a\simeq0.05\pm0.02$ and $0<l<0.5$.}

\begin{figure*}[t!]
	\begin{center}
		a.
		\includegraphics[width=0.45\linewidth]{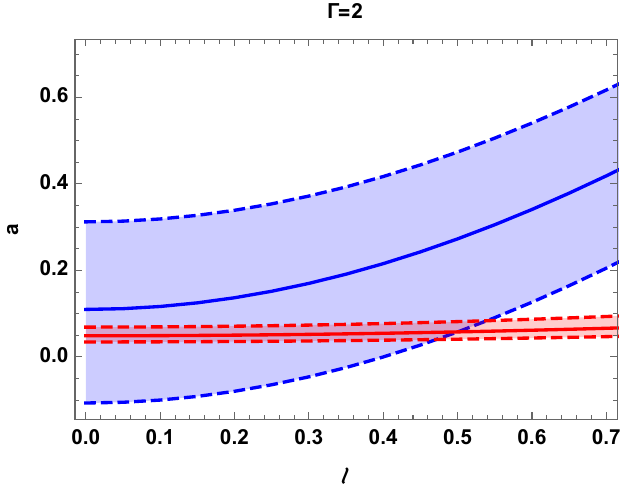}
		b.
		\includegraphics[width=0.45\linewidth]{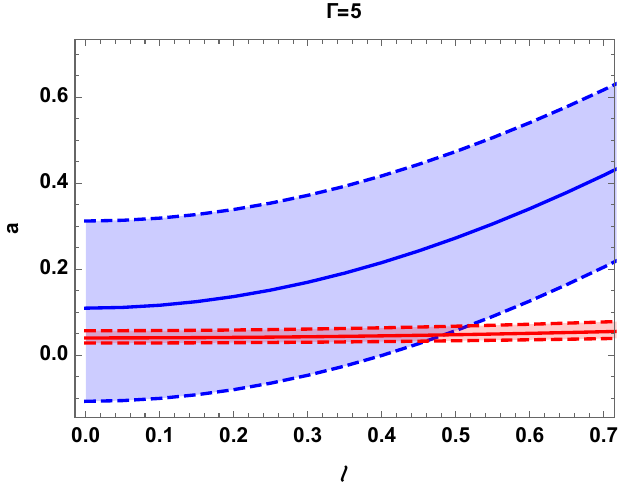}
	\end{center}
	\caption{\textcolor{black}{A0620-00. The blue solid curve indicates the values of the spin parameter $a$ and of the NUT parameter $l$ that reproduce $\eta = 0.061$ (best-fit value in Tab.~\ref{Table1}). The blue dashed curves are for $\eta = 0.054$ and $\eta = 0.070$ (lower and upper constraints in Tab.~\ref{Table1}). The observed thermal spectrum of A0620-00 is compatible with all Kerr-Taub-NUT spacetimes in the shaded region. Red colored regions are for the matching values of the spacetime parameters $a$ and $l$ reproducing the same jet power of the source. The red solid line is for the central value of the jet power while the dashed lines correspond to the case of error of $0.3 \, dex$. Left panel is for the Lorentz factor $\Gamma=2$ and the right one for $\Gamma=5$. Intersection of the blue and red regions describes both observational constraints when the Lorentz factor is $\Gamma=2$ (left panel) and $\Gamma=5$ (right panel). For more detailed information please read the main text. \label{1Je}}}
\end{figure*}

\item {\bf Source H1743-322}. 
The spin of the black hole has been measured by using the CFM in Ref.~ \cite{ref100}: the value of the spin parameter is $0.2\pm0.3$ at the 68\% of confidence level and $-0.3<a<0.7$ at the 90\% of confidence level. The radiative efficiency of the source is thus $0.065^{+0.017}_{-0.011}$  at 68\% CL.
In Fig.~\ref{2Je}, the blue regions represent our constraints on the Kerr-Taub-NUT spacetime parameters to explain the observed radiative efficiency of the source H1743-322. For $l = 0$ (Kerr black hole), the spin parameter would be $a\simeq0.2$. In the Kerr-Taub-NUT spacetime, the spin parameter may be up to $a\simeq0.8$ for $l=1$. The two dashed blue curves are the boundary of the region in the parameter space allowed by observations.
With the red regions in Fig.~\ref{2Je} we have shown results for the jet power of the object H1743-322. It is shown that with the increase of the NUT parameter from $l=0$ up to $l=1$ the corresponding range of the spin parameter to explain the observed jet power within the error bars considerably increased from $a\simeq0.35\pm0.1$ to $a\simeq0.55\pm0.2$, respectively, for the case $\Gamma=2$ in the left panel. Meanwhile, for the case of $\Gamma=5$ this change slightly differs from the case $\Gamma=2$ as can be seen from the right panel. It is also noticeable that, theoretically, the whole range of the NUT parameter of the Kerr-Taub-NUT spacetime is efficient to explain the observed jet power of the source which is apparently shown in the figure.
Similarly to the source A0620-00 we see that in both cases (i.e. $\Gamma=2$ and $\Gamma=5$) the results look almost identical. It is clearly demonstrated that the source H1743-322 described by the Kerr-Taub-NUT spacetime has a good chance to be selected as a source with this spacetime geometry. This is because either in the left or in the right panel one can see the intersection of the central values of the observed radiative efficiency and the jet power. This, in turn, allows taking the corresponding points, $a\simeq0.4$ and $l\simeq0.6$ to be the favorable parameters of this source which can simultaneously explain the values of both observable quantities. One can also notice that the entire red shaded region with the NUT parameter up to $l\simeq1$ can be well used to explain both observational constraints within the error bars using the Kerr-Taub-NUT spacetime.

\begin{figure*}[t!]
	\begin{center}
		a.
		\includegraphics[width=0.45\linewidth]{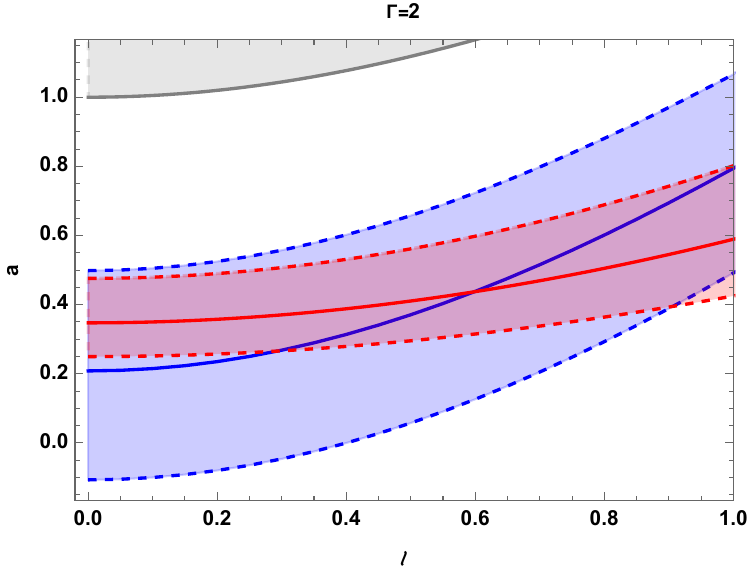}
		b.
		\includegraphics[width=0.45\linewidth]{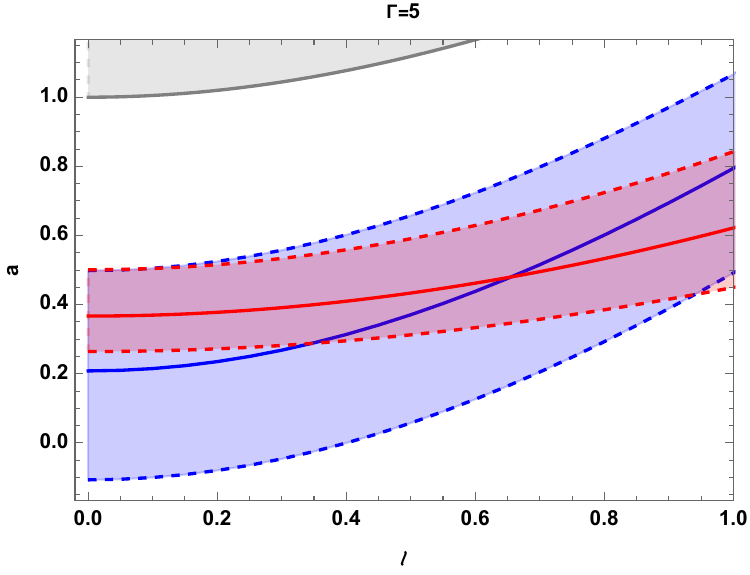}
	\end{center}
	\caption{\textcolor{black}{H1743-322. The blue shaded region marks the values of $a$ and $l$ compatible with the observed spectrum of the source. The gray shaded region shows the case when there is no black hole. As in the previous case the figure (a) is for $\Gamma=2$ and the figure (b) when $\Gamma=5$ and with the same definitions of the lines and regions. From both figures, it comes out that the favorable values of the spin and NUT parameter of the central source are $a\simeq0.4$ and $l\simeq0.6$, respectively.  \label{2Je}}}
\end{figure*}

\item {\bf Source XTE J1550-564}. 
The spin measurement obtained in \cite{steiner2011spin} shows the values $a \simeq 0.34\pm 0.24$ at 68\% CL.
Fig.~\ref{3Je} shows the constraints on $a$ and $l$ when the spacetime metric of its black hole is described by the Kerr-Taub-NUT solution. The spin parameter measurement $a\simeq0.34\pm0.24$ for the case of a Kerr black hole ($l=0$) to explain the radiative efficiency of the source can become $a=1$ when $l\simeq0.85$, and we can have already $a>1$ for $l=1$. 
One can also see that the values of the parameters $a$ and $l$ to describe the observed $P_{jet}$ (red regions) within the error bars are almost the same in both cases namely, in the cases $\Gamma=2$ and $\Gamma=5$. The upper dashed red edge corresponding to the upper value of the jet power with the error of $0.3 \, dex$ comes very close to the extreme spin parameter of the pure Kerr spacetime at $l=1$ while the red solid line providing the central value of $P_{jet}$ given in the tab.~\ref{Table2} corresponds to the object with an intermediate spin parameter with $a\simeq0.7$ at $l=1$. So, the result indicates that the source XTE J1550-564 believed to be the Kerr black hole with the spin parameter $a\simeq0.42\pm0.11$ can also be explained by the Kerr-Taub-NUT spacetime with the spin and NUT parameters corresponding to the shaded regions in Fig.~\ref{3Je} producing the same observed jet power within the error bars.
We see very similar behavior of the regions presented in both panels of Fig.~\ref{3Je} as was in the case of the previous source. The result can be explained in the same way as for the source H1743-322. Indeed, we see the cross point of the two central blue and red solid lines at $a\simeq0.5$ and $l\simeq0.45$ being the favorable values of these parameters to describe the observed $\eta$ and $P_{jet}$. One can also state from the figure that most part of the red shaded region in the panels corresponding to the observed jet power is applicable to explain both observational constraints in the given error bars.

\begin{figure*}[t!]
	\begin{center}
		a.
		\includegraphics[width=0.45\linewidth]{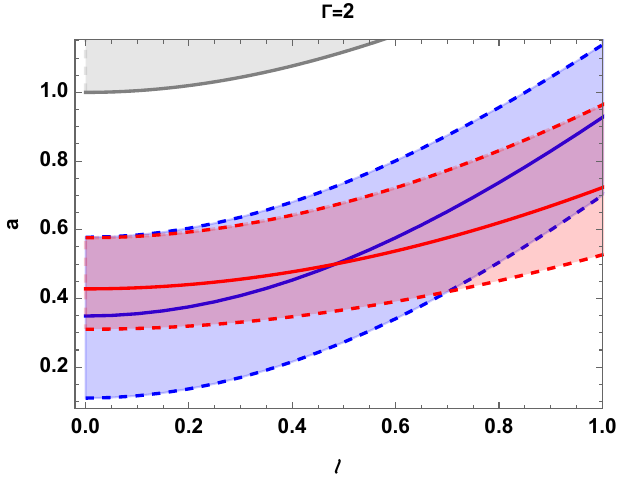}
		b.
		\includegraphics[width=0.45\linewidth]{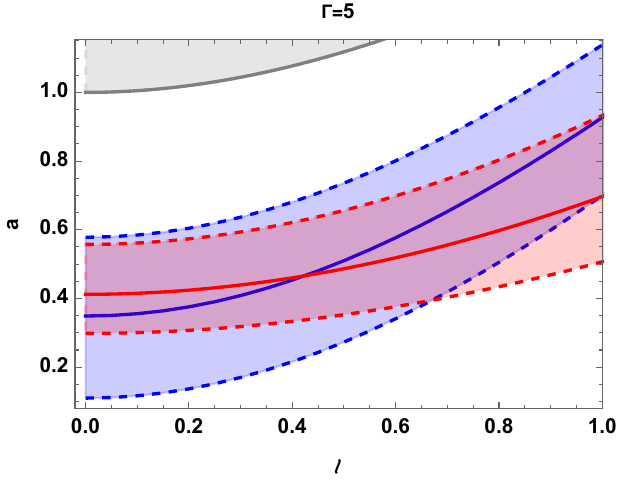}
	\end{center}
	\caption{\textcolor{black}{XTE J1550-564. The explanation of the graphs is similar to the previous case. Please see the corresponding text for proper description.\label{3Je}}}
\end{figure*}

\item {\bf Source GRS 1124-683}. 
Using the CFM and assuming the Kerr metric, the measurement of the spin parameter of the black hole is $0.63^{+0.16}_{-0.19}$  at 68\% CL~\cite{ref109}. 
From such a measurement reported in the literature, we deduce the constraints on $a$ and $l$ shown with blue regions in Fig.~\ref{4Je}. We see that the blue solid line grows up to the values of the spin and NUT parameters of Kerr-Taub-NUT metric $a=1$ and $l\simeq0.8$, respectively. The upper dashed line, which is for the case $\eta=0.095+0.025$, starts from the point $a\simeq0.79$ at $l=0$ and increases up to $a=1.3$ at $l=1$. The lower boundary starts from the point $a\simeq0.44$ at $l=0$ and goes up to $a\simeq1$ at $l=1$.
As for the observed jet power of the source given with red colored regions we see that the cases for the Lorentz factor $\Gamma=2$ in the left and $\Gamma=5$ in the right differ from each other considerably. In the left panel we see that starting from the range $a\simeq0.26^{+0.11}_{-0.08}$ corresponding to the Kerr case the spin parameter increases up to $a\simeq0.45^{+0.2}_{-0.13}$ for the case of $l=1$. In the right panel, however, one can see that for $\Gamma=5$ the starting range corresponds to $a\simeq0.57^{+0.17}_{-0.15}$ in the absence of the NUT parameter and exceeds the extreme rotation of the Kerr BH for the upper error bar when $l=1$. For the central value of the jet power, it grows up to $a\simeq0.95$  when the NUT parameter is taken to be  $l=1$. The figure (b) shows the obvious difference with respect to the case $\Gamma=2$ in the left panel. Generally speaking, in the right panel the upper values of the NUT parameter are spinning up the compact object until rapidly rotating black hole case.
When it comes to the unification of the two observational constraints one can see the big difference in the regions presented in the left and right panels of the figure. For the Lorentz factor $\Gamma=2$ we see that the observed jet power and the radiative efficiency of the source presented in Tab.~\ref{Table1} and Tab.~\ref{Table2} can not be theoretically explained at the same time by the Kerr-Taub-NUT spacetime with the matching values of the parameters $a$ and $l$ since there are no intersecting regions in the left panel of Fig.~\ref{4Je}. One can however see from the right panel of the figure that for the value of the Lorentz factor $\Gamma=5$ there is a region where shaded regions of the two observational constraints intersect with each other. This in turn allows taking the points within this region to be the ones that can explain both, the jet power and the radiative efficiency of the source. Based on the idea that the source is described by the Kerr-Taub-NUT spacetime one can assume that this source corresponds to the object  with a relatively high Lorentz factor that emits radio jets.

\begin{figure*}[t!]
	\begin{center}
		a.
		\includegraphics[width=0.45\linewidth]{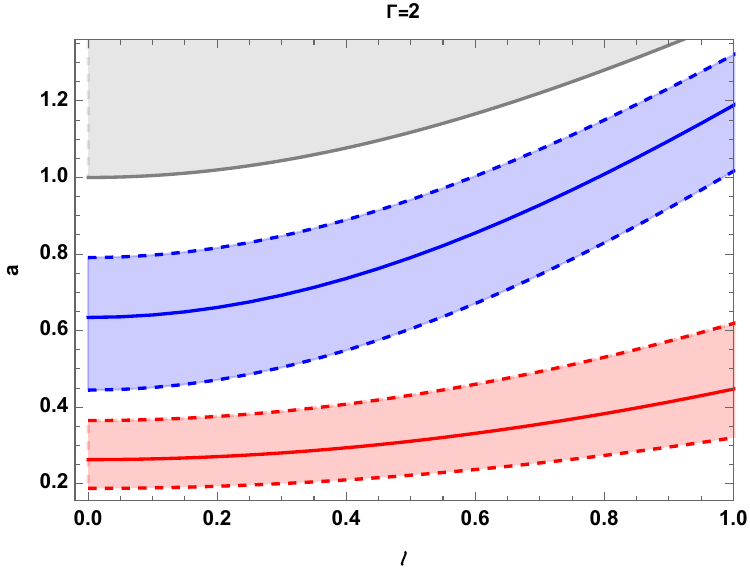}
		b.
		\includegraphics[width=0.45\linewidth]{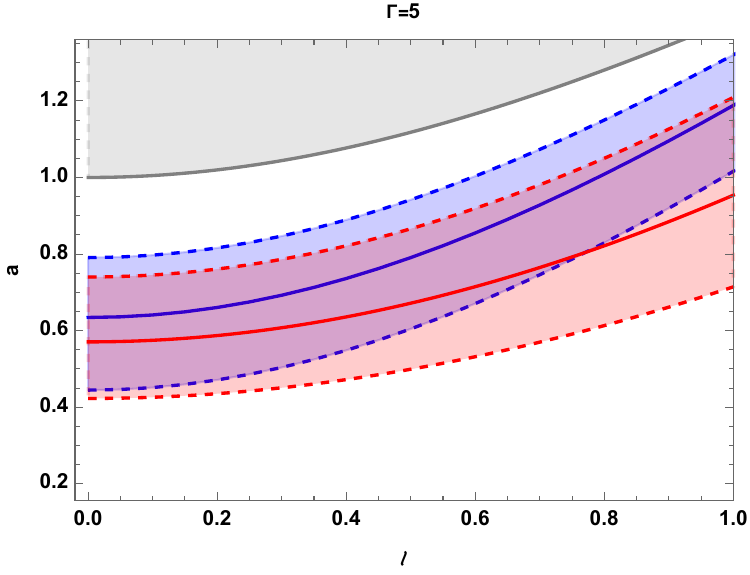}
	\end{center}
	\caption{\textcolor{black}{GRS 1124-683. In the left panel, we see that the Kerr-Taub-NUT spacetime for the Lorentz factor $\Gamma=2$ is unable to explain the observed jet power and the radiative efficiency of the source at the same time. However, in the right panel, we see the intersection of the regions when $\Gamma=5$. For more detailed discussion, see the main text. \label{4Je}}}
\end{figure*}

\item {\bf Source GRO J1655-40}. 
The spin measured by CFM shows $a\simeq 0.7\pm 0.1$ at 68\% CL ~\cite{Shafee06}.
The blue regions in Fig.~\ref{5Je} show the constraints on $a$ and $l$ for observed radiative efficiency when we assume that the spacetime around the black hole is described by the Kerr-Taub-NUT metric. The value of the NUT parameter that can explain the observed $\eta$ is $l\simeq0.8$ when the rotation parameter of the compact object takes the maximum value for a Kerr black hole. For higher NUT parameter we have faster spin of the source.
From the left panel of Fig.~\ref{5Je}, it is seen that theoretically, the Kerr-Taub-NUT spacetime can describe the observed jet power of the source GRO J1655-40  for the entire range of $0<l<1$ for the Lorentz factor $\Gamma=2$. However, from the right panel where $\Gamma=5$ one can see that it is not so when $l>0.8$ where the central solid line crosses the upper error bar. One can state that the Kerr-Taub-NUT spacetime becomes less effective to explain the observed jet power of the source GRO J1655-40 for relatively bigger values of the NUT parameter and bigger Lorentz factor.
Now let us talk about the unification of the two constraints. Having evaluated for the cases $\Gamma=2$ and $\Gamma=5$ the regions experience considerable differences in the left and right panels, respectively. On the left panel, we see that the central value of the jet power is above the upper error bar of the radiative efficiency profile while the central value of the latter is inside of the jet power profile. We see much difference in the right panel where the intersection of the shaded regions is even smaller compared to the case $\Gamma=2$. Theoretically, from the values of the radiative efficiency presented in Tab.~\ref{Table1}  and the jet power in Tab.~\ref{Table2} one can state that the intersection of the regions can be used to explain the observational constraints within the error bars as shown in the figure.

\begin{figure*}[t!]
	\begin{center}
		a.
		\includegraphics[width=0.45\linewidth]{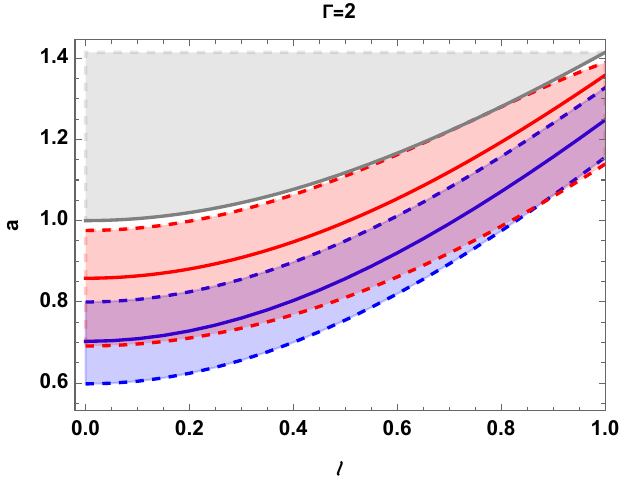}
		b.
		\includegraphics[width=0.45\linewidth]{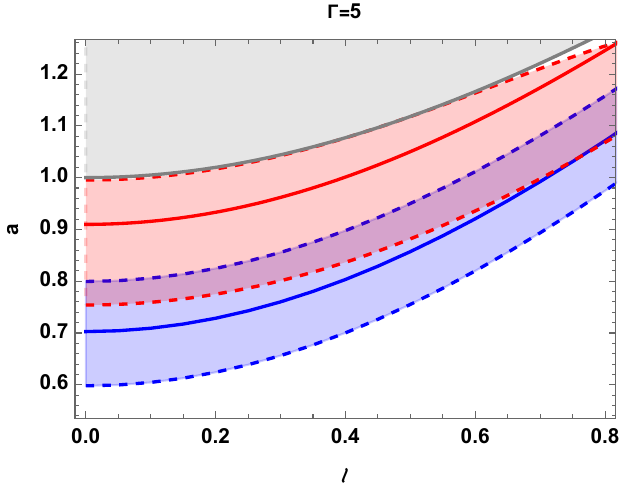}
	\end{center}
	\caption{\textcolor{black}{GRO J1655-40. Panels corresponding to $\Gamma=2$ (left) and $\Gamma=5$ (right) show that the Kerr-Taub-NUT spacetime can be used to explain the observed $\eta$ and at the same time $P_{jet}$. More discussion can be found in the main text.\label{5Je}}}
\end{figure*}

\item {\bf Source GRS~1915+105}. 
In \cite{ref118}, the measurement for the spin is $a > 0.98$ at 68\% CL.
The blue regions in Fig.~\ref{6Je} show the constraints on $a$ and $l$ for the observed radiative efficiency when we assume that GRS~1915+105 hosts a Kerr-Taub-NUT black hole. We see that for this object the largest value of the NUT parameter cannot exceed $l\simeq0.8$ to describe a black hole since we have the intersection of the blue dashed line with the gray shaded region that does not correspond to a black hole.
The behavior of the parameters explaining the observed jet power (red regions) looks similar in both values of the Lorentz factor. From the left and right panels it is apparent that the central value of $P_{jet}$ reaches the extreme rotation of the Kerr BH when the NUT parameter gets closer to $l\simeq0.8$ for both the Lorentz factors $\Gamma=2$ and $\Gamma=5$.
It is worth noting here that, for the pure Kerr spacetime the expression \eqref{PBZ} works very well for spins up to $a\simeq0.95$ and for even higher spins one needs to take higher order terms of $\Omega_H$ \cite{ref89}. This is because in the Kerr spacetime a spin that is close to 1 corresponds to the extremely rotating case. However, as we discussed previously, in the Kerr-Taub-NUT spacetime this is not the case and, in principle, the spin of a black hole can take arbitrarily high values depending on the value of the NUT parameter ($|a| \le \sqrt{1+l^2}$). To ensure we have taken into account even higher orders of $\Omega_H$ which is $P_{BZ}=k \Phi_{tot}^2 (\Omega_H^2\ + \alpha \Omega_H^4 + \beta \Omega_H^6)$ where $\alpha\simeq 1.38$ and $\beta\simeq -9.2$ according to \cite{ref89}. We have found that corrections due to these additional terms are negligible.
From the left and right panels, it is clearly seen that there is no region where the shaded regions of the two observational constraints can intersect with each other. This, in turn, indicates that the observed radiative efficiency and the jet power of the source given in Tab.~\ref{Table1} and Tab.~\ref{Table2}, respectively, cannot be theoretically explained by using the Kerr-Taub-NUT spacetime geometry at the same time. We see from the figure that the region that can explain the radiative efficiency of the source corresponds to the rapid rotation of the source for small values of gravitomagnetic charge while the region to explain the jet power of the source is considerably below the former in the same ranges of the NUT parameter.

\begin{figure*}[t!]
	\begin{center}
		a.
		\includegraphics[width=0.45\linewidth]{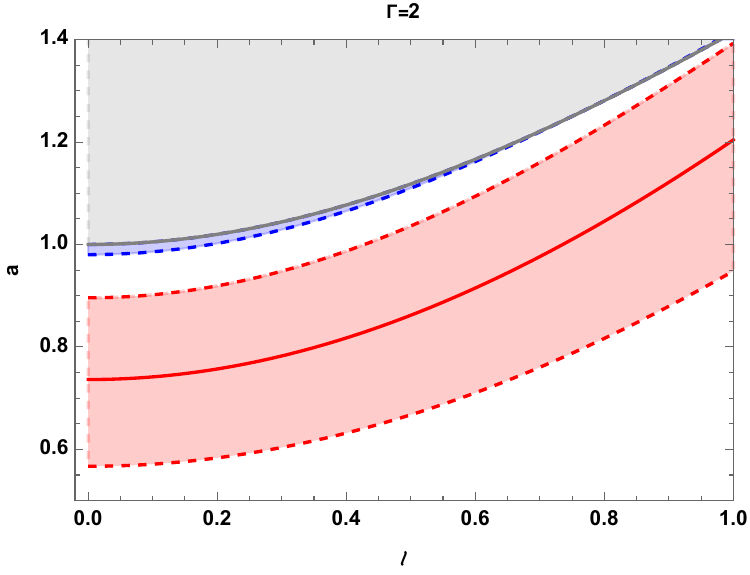}
		b.
		\includegraphics[width=0.45\linewidth]{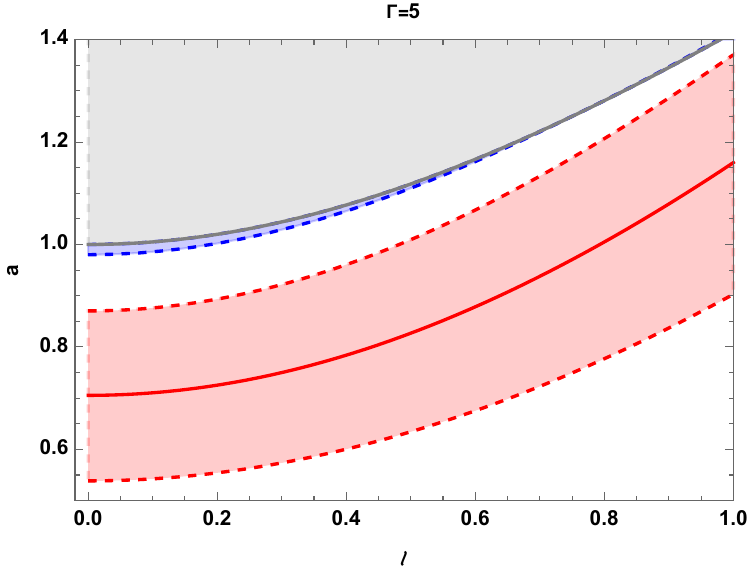}
	\end{center}
	\caption{GRS1915+105. Results in the left and right panels show that the Kerr-Taub-NUT spacetime faces some difficulties to explain both observational constraints. For more discussion see the corresponding text. \label{6Je}}
\end{figure*}

\end{itemize}

It is worth noting that the estimates of the NUT parameter $l$ and of the rotational parameter $a$ are correlated and, eventually, completely degenerate. A higher value of the NUT parameter requires a higher value of the spin of the source to reproduce the radiative efficiency presented in Tab.~\ref{Table1}.
Based on the results in this section one can state that the gravitational sources A0620-00, H1743-322, and XTE J1550-564 have good chances to be described by the Kerr-Taub-NUT spacetime for both values of the Lorentz factor as was discussed above. For the sources GRS1124-683 and GRO J1655-40 it is also possible to explain the two observational constraints at the same time by the Kerr-Taub-NUT spacetime but it is favorable to take the corresponding Lorentz factors to be $\Gamma=5$ for GRS1124-683 and $\Gamma=2$ for GRO J1655-40. On the other hand, the results for the source GRS1915+105 indicate that the chosen spacetime metric is not applicable to explain both observational constraints simultaneously since as was demonstrated above there is no intersection of the regions corresponding to these observational constraints.

\section{Conclusion \label{section4.6}}
In this work, we have applied the Kerr-Taub-NUT metric to describe spacetime geometry around the known sources A0620-00, H1743-322, XTE J1550-564, GRS1124-683, GRO J1655-40, and GRS1915+105. We have briefly introduced the properties of the spacetime that has two extra parameters besides the mass such as spin and NUT parameters. From the idea that if the source is described by the Kerr-Taub-NUT spacetime then the observed radiative efficiency of the black hole source could be interpreted by the corresponding values of the extra spacetime parameters, we have shown the regions within the error bars for these extra parameters of each source. Results have shown that in all cases the range of the spin parameter to describe the observed radiative efficiency shifts upward with the increase of the NUT parameter.

Next, we have applied a similar idea to the jet power of the source to see if the Kerr-Taub-NUT spacetime is able to explain the observed jet power of the selected known sources in some range of the spacetime parameters $a$ and $l$. Our analyses have confirmed that, in principle, the answer to this question is positive. We have demonstrated that one can explain the jet power of the black hole sources by the entire values of the NUT parameter and for bigger values of this parameter we have seen that the corresponding range of the spin parameter goes up similarly to what happened in the case of radiative efficiency of the sources. For the black hole candidates XTE J1550-564, A0620-00, GRS1915+105, and H1743-322 it has been demonstrated that the regions look quite similar to each other for the values of the Lorentz factor $\Gamma=2$ and $\Gamma=5$ while for the source GRS1124-683 and GRO J1655-40 we have found considerable differences. It has been also shown that corresponding to the non-rapidly rotating black hole in the case of the Kerr spacetime (when $l=0$) the sources may become rapidly rotating Kerr-Taub-NUT black holes with the increase of the NUT parameter to explain the observed jet power.

Then, we have unified the results obtained in the last two sections to see if the Kerr-Taub-NUT spacetime is able to explain at the same time both observational quantities. Results presented have shown that the Kerr-Taub-NUT spacetime is well applicable for the sources A0620-00, H1743-322, and XTE J1550-564 to explain both observational phenomena at the same time. For the black hole candidates GRO J1655-40 and GRS1124-683, however, we have seen that the situation changes with the change of the Lorentz factor. From the unification of the results of GRS1915+105 however, we have found that the Kerr-Taub-NUT spacetime gives the regions for the two observational quantities that do not intersect with each other. This in turn indicates that the chosen spacetime is not applicable to reproduce simultaneously the values of radiative efficiency and the jet power of the source presented in Tab.~\ref{Table1} and Tab.~\ref{Table2}. However, one can see that the Kerr spacetime is also not able to simultaneously explain the two observations and this indicates that an additional analysis is requested to explain observational properties of this object. One should also mention that our results assume the conjecture proposed in refs. \cite{ref84} and \cite{ref91}, but some authors criticize these results (see e.g. \cite{Russell2013ws}). For the moment, we cannot say which group is right and which is wrong, as the number of sources is low. In the future, with more data and more precise measurements, we can confirm if the correlation proposed by Narayan and McClintock is correct or not.

\begin{acknowledgments}
	
	This research is supported  by Grants F-FA-2021-432 and MRB-2021-527 of the Uzbekistan Ministry for Innovative Development and by the Abdus Salam International
Centre for Theoretical Physics under the Grant No. OEA-NT-01. 
C.B. acknowledges support from National Natural Science Foundation of China (NSFC), Grant No. 11973019, Natural Science Foundation of Shanghai, Grant No. 22ZR1403400, Shanghai Municipal Education Commission, Grant No. 2019-01-07-00-07-E00035, and Fudan University, Grant No. JIH1512604.

\end{acknowledgments}

\bibliographystyle{apsrev4-1}
\bibliography{gravreferences}

\begin{thebibliography}{59}%
\makeatletter
\providecommand \@ifxundefined [1]{%
 \@ifx{#1\undefined}
}%
\providecommand \@ifnum [1]{%
 \ifnum #1\expandafter \@firstoftwo
 \else \expandafter \@secondoftwo
 \fi
}%
\providecommand \@ifx [1]{%
 \ifx #1\expandafter \@firstoftwo
 \else \expandafter \@secondoftwo
 \fi
}%
\providecommand \natexlab [1]{#1}%
\providecommand \enquote  [1]{``#1''}%
\providecommand \bibnamefont  [1]{#1}%
\providecommand \bibfnamefont [1]{#1}%
\providecommand \citenamefont [1]{#1}%
\providecommand \href@noop [0]{\@secondoftwo}%
\providecommand \href [0]{\begingroup \@sanitize@url \@href}%
\providecommand \@href[1]{\@@startlink{#1}\@@href}%
\providecommand \@@href[1]{\endgroup#1\@@endlink}%
\providecommand \@sanitize@url [0]{\catcode `\\12\catcode `\$12\catcode
  `\&12\catcode `\#12\catcode `\^12\catcode `\_12\catcode `\%12\relax}%
\providecommand \@@startlink[1]{}%
\providecommand \@@endlink[0]{}%
\providecommand \url  [0]{\begingroup\@sanitize@url \@url }%
\providecommand \@url [1]{\endgroup\@href {#1}{\urlprefix }}%
\providecommand \urlprefix  [0]{URL }%
\providecommand \Eprint [0]{\href }%
\providecommand \doibase [0]{http://dx.doi.org/}%
\providecommand \selectlanguage [0]{\@gobble}%
\providecommand \bibinfo  [0]{\@secondoftwo}%
\providecommand \bibfield  [0]{\@secondoftwo}%
\providecommand \translation [1]{[#1]}%
\providecommand \BibitemOpen [0]{}%
\providecommand \bibitemStop [0]{}%
\providecommand \bibitemNoStop [0]{.\EOS\space}%
\providecommand \EOS [0]{\spacefactor3000\relax}%
\providecommand \BibitemShut  [1]{\csname bibitem#1\endcsname}%
\let\auto@bib@innerbib\@empty
\bibitem [{\citenamefont {{Miller}}(1973)}]{Miller73}%
  \BibitemOpen
  \bibfield  {author} {\bibinfo {author} {\bibfnamefont {J.~G.}\ \bibnamefont
  {{Miller}}},\ }\href {\doibase 10.1063/1.1666343} {\bibfield  {journal}
  {\bibinfo  {journal} {Journal of Mathematical Physics}\ }\textbf {\bibinfo
  {volume} {14}},\ \bibinfo {pages} {486} (\bibinfo {year} {1973})}\BibitemShut
  {NoStop}%
\bibitem [{\citenamefont {Hackmann}\ and\ \citenamefont
  {Lammerzahl}(2012)}]{ref25}%
  \BibitemOpen
  \bibfield  {author} {\bibinfo {author} {\bibfnamefont {E.}~\bibnamefont
  {Hackmann}}\ and\ \bibinfo {author} {\bibfnamefont {C.}~\bibnamefont
  {Lammerzahl}},\ }\href {\doibase 10.1103/PhysRevD.85.044049} {\bibfield
  {journal} {\bibinfo  {journal} {Phys. Rev. D}\ }\textbf {\bibinfo {volume}
  {85}},\ \bibinfo {pages} {044049} (\bibinfo {year} {2012})},\ \Eprint
  {http://arxiv.org/abs/1107.5250} {arXiv:1107.5250 [gr-qc]} \BibitemShut
  {NoStop}%
\bibitem [{\citenamefont {Dadhich}\ and\ \citenamefont
  {Turakulov}(2002)}]{Dadhich:2001sz}%
  \BibitemOpen
  \bibfield  {author} {\bibinfo {author} {\bibfnamefont {N.}~\bibnamefont
  {Dadhich}}\ and\ \bibinfo {author} {\bibfnamefont {Z.~Y.}\ \bibnamefont
  {Turakulov}},\ }\href {\doibase 10.1088/0264-9381/19/11/301} {\bibfield
  {journal} {\bibinfo  {journal} {Class. Quant. Grav.}\ }\textbf {\bibinfo
  {volume} {19}},\ \bibinfo {pages} {2765} (\bibinfo {year} {2002})},\ \Eprint
  {http://arxiv.org/abs/gr-qc/0112031} {arXiv:gr-qc/0112031} \BibitemShut
  {NoStop}%
\bibitem [{\citenamefont {{Chakraborty}}\ and\ \citenamefont
  {{Bhattacharyya}}(2018)}]{ref0}%
  \BibitemOpen
  \bibfield  {author} {\bibinfo {author} {\bibfnamefont {C.}~\bibnamefont
  {{Chakraborty}}}\ and\ \bibinfo {author} {\bibfnamefont {S.}~\bibnamefont
  {{Bhattacharyya}}},\ }\href {\doibase 10.1103/PhysRevD.98.043021} {\bibfield
  {journal} {\bibinfo  {journal} {\prd}\ }\textbf {\bibinfo {volume} {98}},\
  \bibinfo {eid} {043021} (\bibinfo {year} {2018})},\ \Eprint
  {http://arxiv.org/abs/1712.01156} {arXiv:1712.01156 [astro-ph.HE]}
  \BibitemShut {NoStop}%
\bibitem [{\citenamefont {{Hakimov}}\ \emph {et~al.}(2017)\citenamefont
  {{Hakimov}}, \citenamefont {{Abdujabbarov}},\ and\ \citenamefont
  {{Narzilloev}}}]{Hakimov17}%
  \BibitemOpen
  \bibfield  {author} {\bibinfo {author} {\bibfnamefont {A.}~\bibnamefont
  {{Hakimov}}}, \bibinfo {author} {\bibfnamefont {A.}~\bibnamefont
  {{Abdujabbarov}}}, \ and\ \bibinfo {author} {\bibfnamefont {B.}~\bibnamefont
  {{Narzilloev}}},\ }\href {\doibase 10.1142/S0217751X17501160} {\bibfield
  {journal} {\bibinfo  {journal} {International Journal of Modern Physics A}\
  }\textbf {\bibinfo {volume} {32}},\ \bibinfo {eid} {1750116} (\bibinfo {year}
  {2017})}\BibitemShut {NoStop}%
\bibitem [{\citenamefont {Narzilloev}\ and\ \citenamefont
  {Ahmedov}(2022)}]{Narzilloev22a}%
  \BibitemOpen
  \bibfield  {author} {\bibinfo {author} {\bibfnamefont {B.}~\bibnamefont
  {Narzilloev}}\ and\ \bibinfo {author} {\bibfnamefont {B.}~\bibnamefont
  {Ahmedov}},\ }\href {\doibase 10.3390/sym14091765} {\bibfield  {journal}
  {\bibinfo  {journal} {Symmetry}\ }\textbf {\bibinfo {volume} {14}} (\bibinfo
  {year} {2022}),\ 10.3390/sym14091765}\BibitemShut {NoStop}%
\bibitem [{\citenamefont {Narzilloev}\ and\ \citenamefont
  {Ahmedov}(2023)}]{Narzilloev22b}%
  \BibitemOpen
  \bibfield  {author} {\bibinfo {author} {\bibfnamefont {B.}~\bibnamefont
  {Narzilloev}}\ and\ \bibinfo {author} {\bibfnamefont {B.}~\bibnamefont
  {Ahmedov}},\ }\href {\doibase 10.1016/j.newast.2022.101922} {\bibfield
  {journal} {\bibinfo  {journal} {New Astron.}\ }\textbf {\bibinfo {volume}
  {98}},\ \bibinfo {pages} {101922} (\bibinfo {year} {2023})}\BibitemShut
  {NoStop}%
\bibitem [{\citenamefont {Narzilloev}\ \emph {et~al.}(2022)\citenamefont
  {Narzilloev}, \citenamefont {Abdujabbarov},\ and\ \citenamefont
  {Hakimov}}]{Narzilloev22c}%
  \BibitemOpen
  \bibfield  {author} {\bibinfo {author} {\bibfnamefont {B.}~\bibnamefont
  {Narzilloev}}, \bibinfo {author} {\bibfnamefont {A.}~\bibnamefont
  {Abdujabbarov}}, \ and\ \bibinfo {author} {\bibfnamefont {A.}~\bibnamefont
  {Hakimov}},\ }\href {\doibase 10.1142/S0217751X22501445} {\bibfield
  {journal} {\bibinfo  {journal} {International Journal of Modern Physics A}\
  }\textbf {\bibinfo {volume} {37}},\ \bibinfo {pages} {2250144} (\bibinfo
  {year} {2022})},\ \Eprint
  {http://arxiv.org/abs/https://doi.org/10.1142/S0217751X22501445}
  {https://doi.org/10.1142/S0217751X22501445} \BibitemShut {NoStop}%
\bibitem [{\citenamefont {{Narzilloev}}\ and\ \citenamefont
  {{Ahmedov}}(2023)}]{Narzilloev23}%
  \BibitemOpen
  \bibfield  {author} {\bibinfo {author} {\bibfnamefont {B.}~\bibnamefont
  {{Narzilloev}}}\ and\ \bibinfo {author} {\bibfnamefont {B.}~\bibnamefont
  {{Ahmedov}}},\ }\href {\doibase 10.3390/sym15020293} {\bibfield  {journal}
  {\bibinfo  {journal} {Symmetry}\ }\textbf {\bibinfo {volume} {15}},\ \bibinfo
  {pages} {293} (\bibinfo {year} {2023})}\BibitemShut {NoStop}%
\bibitem [{\citenamefont {{Mirzaev}}\ \emph {et~al.}(2023)\citenamefont
  {{Mirzaev}}, \citenamefont {{Li}}, \citenamefont {{Narzilloev}},
  \citenamefont {{Hussain}}, \citenamefont {{Abdujabbarov}},\ and\
  \citenamefont {{Ahmedov}}}]{Narzilloev23a}%
  \BibitemOpen
  \bibfield  {author} {\bibinfo {author} {\bibfnamefont {T.}~\bibnamefont
  {{Mirzaev}}}, \bibinfo {author} {\bibfnamefont {S.}~\bibnamefont {{Li}}},
  \bibinfo {author} {\bibfnamefont {B.}~\bibnamefont {{Narzilloev}}}, \bibinfo
  {author} {\bibfnamefont {I.}~\bibnamefont {{Hussain}}}, \bibinfo {author}
  {\bibfnamefont {A.}~\bibnamefont {{Abdujabbarov}}}, \ and\ \bibinfo {author}
  {\bibfnamefont {B.}~\bibnamefont {{Ahmedov}}},\ }\href {\doibase
  10.1140/epjp/s13360-022-03632-4} {\bibfield  {journal} {\bibinfo  {journal}
  {European Physical Journal Plus}\ }\textbf {\bibinfo {volume} {138}},\
  \bibinfo {eid} {47} (\bibinfo {year} {2023})}\BibitemShut {NoStop}%
\bibitem [{\citenamefont {Narzilloev}\ and\ \citenamefont
  {Ahmedov}(2023)}]{Narzilloev2023b}%
  \BibitemOpen
  \bibfield  {author} {\bibinfo {author} {\bibfnamefont {B.}~\bibnamefont
  {Narzilloev}}\ and\ \bibinfo {author} {\bibfnamefont {B.}~\bibnamefont
  {Ahmedov}},\ }\href {\doibase 10.1142/S0217751X23500264} {\bibfield
  {journal} {\bibinfo  {journal} {Int. J. Mod. Phys. A}\ }\textbf {\bibinfo
  {volume} {38}},\ \bibinfo {pages} {2350026} (\bibinfo {year}
  {2023})}\BibitemShut {NoStop}%
\bibitem [{\citenamefont {{Abdulxamidov, Farrux}}\ \emph
  {et~al.}(2023)\citenamefont {{Abdulxamidov, Farrux}}, \citenamefont
  {{Benavides-Gallego, Carlos A.}}, \citenamefont {{Narzilloev, Bakhtiyor}},
  \citenamefont {{Hussain, Ibrar}}, \citenamefont {{Abdujabbarov, Ahmadjon}},
  \citenamefont {{Ahmedov, Bobomurat}},\ and\ \citenamefont {{Xu,
  Haiguang}}}]{Narzilloev2023c}%
  \BibitemOpen
  \bibfield  {author} {\bibinfo {author} {\bibnamefont {{Abdulxamidov,
  Farrux}}}, \bibinfo {author} {\bibnamefont {{Benavides-Gallego, Carlos A.}}},
  \bibinfo {author} {\bibnamefont {{Narzilloev, Bakhtiyor}}}, \bibinfo {author}
  {\bibnamefont {{Hussain, Ibrar}}}, \bibinfo {author} {\bibnamefont
  {{Abdujabbarov, Ahmadjon}}}, \bibinfo {author} {\bibnamefont {{Ahmedov,
  Bobomurat}}}, \ and\ \bibinfo {author} {\bibnamefont {{Xu, Haiguang}}},\
  }\href {\doibase 10.1140/epjp/s13360-023-04283-9} {\bibfield  {journal}
  {\bibinfo  {journal} {Eur. Phys. J. Plus}\ }\textbf {\bibinfo {volume}
  {138}},\ \bibinfo {pages} {635} (\bibinfo {year} {2023})}\BibitemShut
  {NoStop}%
\bibitem [{\citenamefont {Alibekov}\ \emph {et~al.}(2023)\citenamefont
  {Alibekov}, \citenamefont {Narzilloev}, \citenamefont {Abdujabbarov},\ and\
  \citenamefont {Ahmedov}}]{Narzilloev2023d}%
  \BibitemOpen
  \bibfield  {author} {\bibinfo {author} {\bibfnamefont {H.}~\bibnamefont
  {Alibekov}}, \bibinfo {author} {\bibfnamefont {B.}~\bibnamefont
  {Narzilloev}}, \bibinfo {author} {\bibfnamefont {A.}~\bibnamefont
  {Abdujabbarov}}, \ and\ \bibinfo {author} {\bibfnamefont {B.}~\bibnamefont
  {Ahmedov}},\ }\href {\doibase 10.3390/sym15071414} {\bibfield  {journal}
  {\bibinfo  {journal} {Symmetry}\ }\textbf {\bibinfo {volume} {15}} (\bibinfo
  {year} {2023}),\ 10.3390/sym15071414}\BibitemShut {NoStop}%
\bibitem [{\citenamefont {Davlataliev}\ \emph {et~al.}(2023)\citenamefont
  {Davlataliev}, \citenamefont {Narzilloev}, \citenamefont {Hussain},
  \citenamefont {Abdujabbarov},\ and\ \citenamefont
  {Ahmedov}}]{Narzilloev2023e}%
  \BibitemOpen
  \bibfield  {author} {\bibinfo {author} {\bibfnamefont {A.}~\bibnamefont
  {Davlataliev}}, \bibinfo {author} {\bibfnamefont {B.}~\bibnamefont
  {Narzilloev}}, \bibinfo {author} {\bibfnamefont {I.}~\bibnamefont {Hussain}},
  \bibinfo {author} {\bibfnamefont {A.}~\bibnamefont {Abdujabbarov}}, \ and\
  \bibinfo {author} {\bibfnamefont {B.}~\bibnamefont {Ahmedov}},\ }\href
  {\doibase 10.1016/j.dark.2023.101340} {\bibfield  {journal} {\bibinfo
  {journal} {Phys. Dark Univ.}\ }\textbf {\bibinfo {volume} {42}},\ \bibinfo
  {pages} {101340} (\bibinfo {year} {2023})}\BibitemShut {NoStop}%
\bibitem [{\citenamefont {Konoplya}\ \emph {et~al.}(2021)\citenamefont
  {Konoplya}, \citenamefont {Kunz},\ and\ \citenamefont
  {Zhidenko}}]{Konoplya2021qll}%
  \BibitemOpen
  \bibfield  {author} {\bibinfo {author} {\bibfnamefont {R.~A.}\ \bibnamefont
  {Konoplya}}, \bibinfo {author} {\bibfnamefont {J.}~\bibnamefont {Kunz}}, \
  and\ \bibinfo {author} {\bibfnamefont {A.}~\bibnamefont {Zhidenko}},\
  }\href@noop {} {\  (\bibinfo {year} {2021})},\ \Eprint
  {http://arxiv.org/abs/2102.10649} {arXiv:2102.10649 [gr-qc]} \BibitemShut
  {NoStop}%
\bibitem [{\citenamefont {Newman}\ \emph {et~al.}(1963)\citenamefont {Newman},
  \citenamefont {Tamburino},\ and\ \citenamefont {Unti}}]{r4}%
  \BibitemOpen
  \bibfield  {author} {\bibinfo {author} {\bibfnamefont {E.}~\bibnamefont
  {Newman}}, \bibinfo {author} {\bibfnamefont {L.}~\bibnamefont {Tamburino}}, \
  and\ \bibinfo {author} {\bibfnamefont {T.}~\bibnamefont {Unti}},\ }\href
  {\doibase 10.1063/1.1704018} {\bibfield  {journal} {\bibinfo  {journal} {J.
  Math. Phys.}\ }\textbf {\bibinfo {volume} {4}},\ \bibinfo {pages} {915}
  (\bibinfo {year} {1963})}\BibitemShut {NoStop}%
\bibitem [{\citenamefont {Misner}(1963)}]{r5}%
  \BibitemOpen
  \bibfield  {author} {\bibinfo {author} {\bibfnamefont {C.~W.}\ \bibnamefont
  {Misner}},\ }\href {\doibase 10.1063/1.1704019} {\bibfield  {journal}
  {\bibinfo  {journal} {J. Math. Phys.}\ }\textbf {\bibinfo {volume} {4}},\
  \bibinfo {pages} {924} (\bibinfo {year} {1963})}\BibitemShut {NoStop}%
\bibitem [{\citenamefont {Lynden-Bell}\ and\ \citenamefont
  {Nouri-Zonoz}(1998)}]{r6}%
  \BibitemOpen
  \bibfield  {author} {\bibinfo {author} {\bibfnamefont {D.}~\bibnamefont
  {Lynden-Bell}}\ and\ \bibinfo {author} {\bibfnamefont {M.}~\bibnamefont
  {Nouri-Zonoz}},\ }\href {\doibase 10.1103/RevModPhys.70.427} {\bibfield
  {journal} {\bibinfo  {journal} {Rev. Mod. Phys.}\ }\textbf {\bibinfo {volume}
  {70}},\ \bibinfo {pages} {427} (\bibinfo {year} {1998})},\ \Eprint
  {http://arxiv.org/abs/gr-qc/9612049} {arXiv:gr-qc/9612049} \BibitemShut
  {NoStop}%
\bibitem [{\citenamefont {{Demianski}}\ and\ \citenamefont
  {{Newman}}(1966)}]{r9}%
  \BibitemOpen
  \bibfield  {author} {\bibinfo {author} {\bibfnamefont {M.}~\bibnamefont
  {{Demianski}}}\ and\ \bibinfo {author} {\bibfnamefont {E.~T.}\ \bibnamefont
  {{Newman}}},\ }\href@noop {} {\bibfield  {journal} {\bibinfo  {journal}
  {Bulletin de l'Academie Polonaise des Sciences Series des Sciences
  Mathematiques Astronomiques et Physiques}\ }\textbf {\bibinfo {volume}
  {14}},\ \bibinfo {pages} {653} (\bibinfo {year} {1966})}\BibitemShut
  {NoStop}%
\bibitem [{\citenamefont {Dirac}(1931)}]{r10}%
  \BibitemOpen
  \bibfield  {author} {\bibinfo {author} {\bibfnamefont {P.~A.~M.}\
  \bibnamefont {Dirac}},\ }\href {\doibase 10.1098/rspa.1931.0130} {\bibfield
  {journal} {\bibinfo  {journal} {Proc. Roy. Soc. Lond. A}\ }\textbf {\bibinfo
  {volume} {133}},\ \bibinfo {pages} {60} (\bibinfo {year} {1931})}\BibitemShut
  {NoStop}%
\bibitem [{\citenamefont {{Saha}}(1936)}]{r11}%
  \BibitemOpen
  \bibfield  {author} {\bibinfo {author} {\bibfnamefont {M.~N.}\ \bibnamefont
  {{Saha}}},\ }\href@noop {} {\bibfield  {journal} {\bibinfo  {journal} {Indian
  J. Phys.}\ }\textbf {\bibinfo {volume} {10}},\ \bibinfo {pages} {141}
  (\bibinfo {year} {1936})}\BibitemShut {NoStop}%
\bibitem [{\citenamefont {{Bonnor}}(1969)}]{r12}%
  \BibitemOpen
  \bibfield  {author} {\bibinfo {author} {\bibfnamefont {W.~B.}\ \bibnamefont
  {{Bonnor}}},\ }\href {\doibase 10.1017/S0305004100044807} {\bibfield
  {journal} {\bibinfo  {journal} {Proceedings of the Cambridge Philosophical
  Society}\ }\textbf {\bibinfo {volume} {66}},\ \bibinfo {pages} {145}
  (\bibinfo {year} {1969})}\BibitemShut {NoStop}%
\bibitem [{\citenamefont {Ramaswamy}\ and\ \citenamefont {Sen}(1981)}]{r7}%
  \BibitemOpen
  \bibfield  {author} {\bibinfo {author} {\bibfnamefont {S.}~\bibnamefont
  {Ramaswamy}}\ and\ \bibinfo {author} {\bibfnamefont {A.}~\bibnamefont
  {Sen}},\ }\href {\doibase 10.1063/1.524839} {\bibfield  {journal} {\bibinfo
  {journal} {Journal of Mathematical Physics}\ }\textbf {\bibinfo {volume}
  {22}},\ \bibinfo {pages} {2612} (\bibinfo {year} {1981})}\BibitemShut
  {NoStop}%
\bibitem [{\citenamefont {{Dowker}}(1974)}]{r13}%
  \BibitemOpen
  \bibfield  {author} {\bibinfo {author} {\bibfnamefont {J.~S.}\ \bibnamefont
  {{Dowker}}},\ }\href {\doibase 10.1007/BF02451402} {\bibfield  {journal}
  {\bibinfo  {journal} {General Relativity and Gravitation}\ }\textbf {\bibinfo
  {volume} {5}},\ \bibinfo {pages} {603} (\bibinfo {year} {1974})}\BibitemShut
  {NoStop}%
\bibitem [{\citenamefont {Kagramanova}\ \emph {et~al.}(2010)\citenamefont
  {Kagramanova}, \citenamefont {Kunz}, \citenamefont {Hackmann},\ and\
  \citenamefont {L\"ammerzahl}}]{r14}%
  \BibitemOpen
  \bibfield  {author} {\bibinfo {author} {\bibfnamefont {V.}~\bibnamefont
  {Kagramanova}}, \bibinfo {author} {\bibfnamefont {J.}~\bibnamefont {Kunz}},
  \bibinfo {author} {\bibfnamefont {E.}~\bibnamefont {Hackmann}}, \ and\
  \bibinfo {author} {\bibfnamefont {C.}~\bibnamefont {L\"ammerzahl}},\ }\href
  {\doibase 10.1103/PhysRevD.81.124044} {\bibfield  {journal} {\bibinfo
  {journal} {Phys. Rev. D}\ }\textbf {\bibinfo {volume} {81}},\ \bibinfo
  {pages} {124044} (\bibinfo {year} {2010})}\BibitemShut {NoStop}%
\bibitem [{\citenamefont {Nouri-Zonoz}\ and\ \citenamefont
  {Lynden-Bell}(1997)}]{r15}%
  \BibitemOpen
  \bibfield  {author} {\bibinfo {author} {\bibfnamefont {M.}~\bibnamefont
  {Nouri-Zonoz}}\ and\ \bibinfo {author} {\bibfnamefont {D.}~\bibnamefont
  {Lynden-Bell}},\ }\href@noop {} {\bibfield  {journal} {\bibinfo  {journal}
  {Mon. Not. Roy. Astron. Soc.}\ }\textbf {\bibinfo {volume} {292}},\ \bibinfo
  {pages} {714} (\bibinfo {year} {1997})},\ \Eprint
  {http://arxiv.org/abs/gr-qc/9812094} {arXiv:gr-qc/9812094} \BibitemShut
  {NoStop}%
\bibitem [{\citenamefont {Rahvar}\ and\ \citenamefont
  {Nouri-Zonoz}(2003)}]{r16}%
  \BibitemOpen
  \bibfield  {author} {\bibinfo {author} {\bibfnamefont {S.}~\bibnamefont
  {Rahvar}}\ and\ \bibinfo {author} {\bibfnamefont {M.}~\bibnamefont
  {Nouri-Zonoz}},\ }\href {\doibase 10.1046/j.1365-8711.2003.06137.x}
  {\bibfield  {journal} {\bibinfo  {journal} {Mon. Not. Roy. Astron. Soc.}\
  }\textbf {\bibinfo {volume} {338}},\ \bibinfo {pages} {926} (\bibinfo {year}
  {2003})},\ \Eprint {http://arxiv.org/abs/astro-ph/0204282}
  {arXiv:astro-ph/0204282} \BibitemShut {NoStop}%
\bibitem [{\citenamefont {{Morozova}}\ \emph {et~al.}(2008)\citenamefont
  {{Morozova}}, \citenamefont {{Ahmedov}},\ and\ \citenamefont
  {{Kagramanova}}}]{r17}%
  \BibitemOpen
  \bibfield  {author} {\bibinfo {author} {\bibfnamefont {V.~S.}\ \bibnamefont
  {{Morozova}}}, \bibinfo {author} {\bibfnamefont {B.~J.}\ \bibnamefont
  {{Ahmedov}}}, \ and\ \bibinfo {author} {\bibfnamefont {V.~G.}\ \bibnamefont
  {{Kagramanova}}},\ }\href {\doibase 10.1086/590322} {\bibfield  {journal}
  {\bibinfo  {journal} {\apj}\ }\textbf {\bibinfo {volume} {684}},\ \bibinfo
  {pages} {1359} (\bibinfo {year} {2008})},\ \Eprint
  {http://arxiv.org/abs/0806.2376} {arXiv:0806.2376 [astro-ph]} \BibitemShut
  {NoStop}%
\bibitem [{\citenamefont {Mirabel}\ and\ \citenamefont
  {Rodriguez}(1999)}]{ref77}%
  \BibitemOpen
  \bibfield  {author} {\bibinfo {author} {\bibfnamefont {I.~F.}\ \bibnamefont
  {Mirabel}}\ and\ \bibinfo {author} {\bibfnamefont {L.~F.}\ \bibnamefont
  {Rodriguez}},\ }\href {\doibase 10.1146/annurev.astro.37.1.409} {\bibfield
  {journal} {\bibinfo  {journal} {Ann. Rev. Astron. Astrophys.}\ }\textbf
  {\bibinfo {volume} {37}},\ \bibinfo {pages} {409} (\bibinfo {year} {1999})},\
  \Eprint {http://arxiv.org/abs/astro-ph/9902062} {arXiv:astro-ph/9902062}
  \BibitemShut {NoStop}%
\bibitem [{\citenamefont {Novikov}\ and\ \citenamefont {Thorne}(1973)}]{ref70}%
  \BibitemOpen
  \bibfield  {author} {\bibinfo {author} {\bibfnamefont {I.~D.}\ \bibnamefont
  {Novikov}}\ and\ \bibinfo {author} {\bibfnamefont {K.~S.}\ \bibnamefont
  {Thorne}},\ }in\ \href@noop {} {\emph {\bibinfo {booktitle} {{Les Houches
  Summer School of Theoretical Physics}: {Black Holes}}}}\ (\bibinfo {year}
  {1973})\ pp.\ \bibinfo {pages} {343--550}\BibitemShut {NoStop}%
\bibitem [{\citenamefont {Bambi}(2012{\natexlab{a}})}]{ref78}%
  \BibitemOpen
  \bibfield  {author} {\bibinfo {author} {\bibfnamefont {C.}~\bibnamefont
  {Bambi}},\ }\href {\doibase 10.1103/PhysRevD.85.043002} {\bibfield  {journal}
  {\bibinfo  {journal} {Phys. Rev. D}\ }\textbf {\bibinfo {volume} {85}},\
  \bibinfo {pages} {043002} (\bibinfo {year} {2012}{\natexlab{a}})},\ \Eprint
  {http://arxiv.org/abs/1201.1638} {arXiv:1201.1638 [gr-qc]} \BibitemShut
  {NoStop}%
\bibitem [{\citenamefont {Narayan}\ and\ \citenamefont
  {McClintock}(2012)}]{ref84}%
  \BibitemOpen
  \bibfield  {author} {\bibinfo {author} {\bibfnamefont {R.}~\bibnamefont
  {Narayan}}\ and\ \bibinfo {author} {\bibfnamefont {J.~E.}\ \bibnamefont
  {McClintock}},\ }\href {\doibase 10.1111/j.1745-3933.2011.01181.x} {\bibfield
   {journal} {\bibinfo  {journal} {Mon. Not. Roy. Astron. Soc.}\ }\textbf
  {\bibinfo {volume} {419}},\ \bibinfo {pages} {L69} (\bibinfo {year}
  {2012})},\ \Eprint {http://arxiv.org/abs/1112.0569} {arXiv:1112.0569
  [astro-ph.HE]} \BibitemShut {NoStop}%
\bibitem [{\citenamefont {Steiner}\ \emph {et~al.}(2013)\citenamefont
  {Steiner}, \citenamefont {McClintock},\ and\ \citenamefont
  {Narayan}}]{ref91}%
  \BibitemOpen
  \bibfield  {author} {\bibinfo {author} {\bibfnamefont {J.~F.}\ \bibnamefont
  {Steiner}}, \bibinfo {author} {\bibfnamefont {J.~E.}\ \bibnamefont
  {McClintock}}, \ and\ \bibinfo {author} {\bibfnamefont {R.}~\bibnamefont
  {Narayan}},\ }\href {\doibase 10.1088/0004-637X/762/2/104} {\bibfield
  {journal} {\bibinfo  {journal} {Astrophys. J.}\ }\textbf {\bibinfo {volume}
  {762}},\ \bibinfo {pages} {104} (\bibinfo {year} {2013})},\ \Eprint
  {http://arxiv.org/abs/1211.5379} {arXiv:1211.5379 [astro-ph.HE]} \BibitemShut
  {NoStop}%
\bibitem [{\citenamefont {Bambi}(2012{\natexlab{b}})}]{PhysRevD86123013}%
  \BibitemOpen
  \bibfield  {author} {\bibinfo {author} {\bibfnamefont {C.}~\bibnamefont
  {Bambi}},\ }\href {\doibase 10.1103/PhysRevD.86.123013} {\bibfield  {journal}
  {\bibinfo  {journal} {Phys. Rev. D}\ }\textbf {\bibinfo {volume} {86}},\
  \bibinfo {pages} {123013} (\bibinfo {year} {2012}{\natexlab{b}})}\BibitemShut
  {NoStop}%
\bibitem [{\citenamefont {{Abdujabbarov}}\ \emph {et~al.}(2008)\citenamefont
  {{Abdujabbarov}}, \citenamefont {{Ahmedov}},\ and\ \citenamefont
  {{Kagramanova}}}]{Abdujabbarov08}%
  \BibitemOpen
  \bibfield  {author} {\bibinfo {author} {\bibfnamefont {A.~A.}\ \bibnamefont
  {{Abdujabbarov}}}, \bibinfo {author} {\bibfnamefont {B.~J.}\ \bibnamefont
  {{Ahmedov}}}, \ and\ \bibinfo {author} {\bibfnamefont {V.~G.}\ \bibnamefont
  {{Kagramanova}}},\ }\href {\doibase 10.1007/s10714-008-0635-3} {\bibfield
  {journal} {\bibinfo  {journal} {General Relativity and Gravitation}\ }\textbf
  {\bibinfo {volume} {40}},\ \bibinfo {pages} {2515} (\bibinfo {year}
  {2008})},\ \Eprint {http://arxiv.org/abs/0802.4349} {arXiv:0802.4349 [gr-qc]}
  \BibitemShut {NoStop}%
\bibitem [{\citenamefont {Page}\ and\ \citenamefont {Thorne}(1974)}]{ref79}%
  \BibitemOpen
  \bibfield  {author} {\bibinfo {author} {\bibfnamefont {D.~N.}\ \bibnamefont
  {Page}}\ and\ \bibinfo {author} {\bibfnamefont {K.~S.}\ \bibnamefont
  {Thorne}},\ }\href {\doibase 10.1086/152990} {\bibfield  {journal} {\bibinfo
  {journal} {Astrophys. J.}\ }\textbf {\bibinfo {volume} {191}},\ \bibinfo
  {pages} {499} (\bibinfo {year} {1974})}\BibitemShut {NoStop}%
\bibitem [{\citenamefont {Bambi}(2017)}]{Bambi17e}%
  \BibitemOpen
  \bibfield  {author} {\bibinfo {author} {\bibfnamefont {C.}~\bibnamefont
  {Bambi}},\ }\href@noop {} {\emph {\bibinfo {title} {Black Holes: A Laboratory
  for Testing Strong Gravity}}}\ (\bibinfo  {publisher} {Springer, Singapore},\
  \bibinfo {year} {2017})\BibitemShut {NoStop}%
\bibitem [{\citenamefont {{Bambi}}\ \emph {et~al.}(2021)\citenamefont
  {{Bambi}}, \citenamefont {{Brenneman}}, \citenamefont {{Dauser}},
  \citenamefont {{Garc{\'\i}a}}, \citenamefont {{Grinberg}}, \citenamefont
  {{Ingram}}, \citenamefont {{Jiang}}, \citenamefont {{Liu}}, \citenamefont
  {{Lohfink}}, \citenamefont {{Marinucci}}, \citenamefont {{Mastroserio}},
  \citenamefont {{Middei}}, \citenamefont {{Nampalliwar}}, \citenamefont
  {{Nied{\'z}wiecki}}, \citenamefont {{Steiner}}, \citenamefont {{Tripathi}},\
  and\ \citenamefont {{Zdziarski}}}]{2021SSRv..217...65B}%
  \BibitemOpen
  \bibfield  {author} {\bibinfo {author} {\bibfnamefont {C.}~\bibnamefont
  {{Bambi}}}, \bibinfo {author} {\bibfnamefont {L.~W.}\ \bibnamefont
  {{Brenneman}}}, \bibinfo {author} {\bibfnamefont {T.}~\bibnamefont
  {{Dauser}}}, \bibinfo {author} {\bibfnamefont {J.~A.}\ \bibnamefont
  {{Garc{\'\i}a}}}, \bibinfo {author} {\bibfnamefont {V.}~\bibnamefont
  {{Grinberg}}}, \bibinfo {author} {\bibfnamefont {A.}~\bibnamefont
  {{Ingram}}}, \bibinfo {author} {\bibfnamefont {J.}~\bibnamefont {{Jiang}}},
  \bibinfo {author} {\bibfnamefont {H.}~\bibnamefont {{Liu}}}, \bibinfo
  {author} {\bibfnamefont {A.~M.}\ \bibnamefont {{Lohfink}}}, \bibinfo {author}
  {\bibfnamefont {A.}~\bibnamefont {{Marinucci}}}, \bibinfo {author}
  {\bibfnamefont {G.}~\bibnamefont {{Mastroserio}}}, \bibinfo {author}
  {\bibfnamefont {R.}~\bibnamefont {{Middei}}}, \bibinfo {author}
  {\bibfnamefont {S.}~\bibnamefont {{Nampalliwar}}}, \bibinfo {author}
  {\bibfnamefont {A.}~\bibnamefont {{Nied{\'z}wiecki}}}, \bibinfo {author}
  {\bibfnamefont {J.~F.}\ \bibnamefont {{Steiner}}}, \bibinfo {author}
  {\bibfnamefont {A.}~\bibnamefont {{Tripathi}}}, \ and\ \bibinfo {author}
  {\bibfnamefont {A.~A.}\ \bibnamefont {{Zdziarski}}},\ }\href {\doibase
  10.1007/s11214-021-00841-8} {\bibfield  {journal} {\bibinfo  {journal} {Space
  Sci. Rev.}\ }\textbf {\bibinfo {volume} {217}},\ \bibinfo {eid} {65}
  (\bibinfo {year} {2021})},\ \Eprint {http://arxiv.org/abs/2011.04792}
  {arXiv:2011.04792 [astro-ph.HE]} \BibitemShut {NoStop}%
\bibitem [{\citenamefont {Zhang}\ \emph {et~al.}(1997)\citenamefont {Zhang},
  \citenamefont {Cui},\ and\ \citenamefont {Chen}}]{Zhang_1997}%
  \BibitemOpen
  \bibfield  {author} {\bibinfo {author} {\bibfnamefont {S.~N.}\ \bibnamefont
  {Zhang}}, \bibinfo {author} {\bibfnamefont {W.}~\bibnamefont {Cui}}, \ and\
  \bibinfo {author} {\bibfnamefont {W.}~\bibnamefont {Chen}},\ }\href {\doibase
  10.1086/310705} {\bibfield  {journal} {\bibinfo  {journal} {The Astrophysical
  Journal}\ }\textbf {\bibinfo {volume} {482}},\ \bibinfo {pages} {L155}
  (\bibinfo {year} {1997})}\BibitemShut {NoStop}%
\bibitem [{\citenamefont {McClintock}\ \emph {et~al.}(2014)\citenamefont
  {McClintock}, \citenamefont {Narayan},\ and\ \citenamefont
  {Steiner}}]{McClintock:2013vwa}%
  \BibitemOpen
  \bibfield  {author} {\bibinfo {author} {\bibfnamefont {J.~E.}\ \bibnamefont
  {McClintock}}, \bibinfo {author} {\bibfnamefont {R.}~\bibnamefont {Narayan}},
  \ and\ \bibinfo {author} {\bibfnamefont {J.~F.}\ \bibnamefont {Steiner}},\
  }\href {\doibase 10.1007/s11214-013-0003-9} {\bibfield  {journal} {\bibinfo
  {journal} {Space Sci. Rev.}\ }\textbf {\bibinfo {volume} {183}},\ \bibinfo
  {pages} {295} (\bibinfo {year} {2014})},\ \Eprint
  {http://arxiv.org/abs/1303.1583} {arXiv:1303.1583 [astro-ph.HE]} \BibitemShut
  {NoStop}%
\bibitem [{\citenamefont {{Kong}}\ \emph {et~al.}(2014)\citenamefont {{Kong}},
  \citenamefont {{Li}},\ and\ \citenamefont {{Bambi}}}]{Kong14}%
  \BibitemOpen
  \bibfield  {author} {\bibinfo {author} {\bibfnamefont {L.}~\bibnamefont
  {{Kong}}}, \bibinfo {author} {\bibfnamefont {Z.}~\bibnamefont {{Li}}}, \ and\
  \bibinfo {author} {\bibfnamefont {C.}~\bibnamefont {{Bambi}}},\ }\href
  {\doibase 10.1088/0004-637X/797/2/78} {\bibfield  {journal} {\bibinfo
  {journal} {Astrophys. J.}\ }\textbf {\bibinfo {volume} {797}},\ \bibinfo
  {eid} {78} (\bibinfo {year} {2014})},\ \Eprint
  {http://arxiv.org/abs/1405.1508} {arXiv:1405.1508 [gr-qc]} \BibitemShut
  {NoStop}%
\bibitem [{\citenamefont {Fender}\ and\ \citenamefont {Belloni}(2004)}]{ref68}%
  \BibitemOpen
  \bibfield  {author} {\bibinfo {author} {\bibfnamefont {R.}~\bibnamefont
  {Fender}}\ and\ \bibinfo {author} {\bibfnamefont {T.}~\bibnamefont
  {Belloni}},\ }\href {\doibase 10.1146/annurev.astro.42.053102.134031}
  {\bibfield  {journal} {\bibinfo  {journal} {Ann. Rev. Astron. Astrophys.}\
  }\textbf {\bibinfo {volume} {42}},\ \bibinfo {pages} {317} (\bibinfo {year}
  {2004})},\ \Eprint {http://arxiv.org/abs/astro-ph/0406483}
  {arXiv:astro-ph/0406483} \BibitemShut {NoStop}%
\bibitem [{\citenamefont {Markoff}\ \emph {et~al.}(2005)\citenamefont
  {Markoff}, \citenamefont {Nowak},\ and\ \citenamefont {Wilms}}]{ref83}%
  \BibitemOpen
  \bibfield  {author} {\bibinfo {author} {\bibfnamefont {S.}~\bibnamefont
  {Markoff}}, \bibinfo {author} {\bibfnamefont {M.~A.}\ \bibnamefont {Nowak}},
  \ and\ \bibinfo {author} {\bibfnamefont {J.}~\bibnamefont {Wilms}},\ }\href
  {\doibase 10.1086/497628} {\bibfield  {journal} {\bibinfo  {journal}
  {Astrophys. J.}\ }\textbf {\bibinfo {volume} {635}},\ \bibinfo {pages} {1203}
  (\bibinfo {year} {2005})},\ \Eprint {http://arxiv.org/abs/astro-ph/0509028}
  {arXiv:astro-ph/0509028} \BibitemShut {NoStop}%
\bibitem [{\citenamefont {{Punsly}}\ and\ \citenamefont
  {{Coroniti}}(1990)}]{ref85}%
  \BibitemOpen
  \bibfield  {author} {\bibinfo {author} {\bibfnamefont {B.}~\bibnamefont
  {{Punsly}}}\ and\ \bibinfo {author} {\bibfnamefont {F.~V.}\ \bibnamefont
  {{Coroniti}}},\ }\href {\doibase 10.1086/168717} {\bibfield  {journal}
  {\bibinfo  {journal} {\apj}\ }\textbf {\bibinfo {volume} {354}},\ \bibinfo
  {pages} {583} (\bibinfo {year} {1990})}\BibitemShut {NoStop}%
\bibitem [{\citenamefont {Koide}(2003)}]{ref87}%
  \BibitemOpen
  \bibfield  {author} {\bibinfo {author} {\bibfnamefont {S.}~\bibnamefont
  {Koide}},\ }\href {\doibase 10.1103/PhysRevD.67.104010} {\bibfield  {journal}
  {\bibinfo  {journal} {Phys. Rev. D}\ }\textbf {\bibinfo {volume} {67}},\
  \bibinfo {pages} {104010} (\bibinfo {year} {2003})}\BibitemShut {NoStop}%
\bibitem [{\citenamefont {Blandford}\ and\ \citenamefont
  {Znajek}(1977)}]{ref69}%
  \BibitemOpen
  \bibfield  {author} {\bibinfo {author} {\bibfnamefont {R.~D.}\ \bibnamefont
  {Blandford}}\ and\ \bibinfo {author} {\bibfnamefont {R.~L.}\ \bibnamefont
  {Znajek}},\ }\href {\doibase 10.1093/mnras/179.3.433} {\bibfield  {journal}
  {\bibinfo  {journal} {Mon. Not. Roy. Astron. Soc.}\ }\textbf {\bibinfo
  {volume} {179}},\ \bibinfo {pages} {433} (\bibinfo {year}
  {1977})}\BibitemShut {NoStop}%
\bibitem [{\citenamefont {Tchekhovskoy}\ \emph {et~al.}(2010)\citenamefont
  {Tchekhovskoy}, \citenamefont {Narayan},\ and\ \citenamefont
  {McKinney}}]{ref89}%
  \BibitemOpen
  \bibfield  {author} {\bibinfo {author} {\bibfnamefont {A.}~\bibnamefont
  {Tchekhovskoy}}, \bibinfo {author} {\bibfnamefont {R.}~\bibnamefont
  {Narayan}}, \ and\ \bibinfo {author} {\bibfnamefont {J.~C.}\ \bibnamefont
  {McKinney}},\ }\href {\doibase 10.1088/0004-637X/711/1/50} {\bibfield
  {journal} {\bibinfo  {journal} {Astrophys. J.}\ }\textbf {\bibinfo {volume}
  {711}},\ \bibinfo {pages} {50} (\bibinfo {year} {2010})},\ \Eprint
  {http://arxiv.org/abs/0911.2228} {arXiv:0911.2228 [astro-ph.HE]} \BibitemShut
  {NoStop}%
\bibitem [{\citenamefont {Camilloni}\ \emph {et~al.}(2022)\citenamefont
  {Camilloni}, \citenamefont {Dias}, \citenamefont {Grignani}, \citenamefont
  {Harmark}, \citenamefont {Oliveri}, \citenamefont {Orselli}, \citenamefont
  {Placidi},\ and\ \citenamefont {Santos}}]{Camilloni:2022kmx}%
  \BibitemOpen
  \bibfield  {author} {\bibinfo {author} {\bibfnamefont {F.}~\bibnamefont
  {Camilloni}}, \bibinfo {author} {\bibfnamefont {O.~J.~C.}\ \bibnamefont
  {Dias}}, \bibinfo {author} {\bibfnamefont {G.}~\bibnamefont {Grignani}},
  \bibinfo {author} {\bibfnamefont {T.}~\bibnamefont {Harmark}}, \bibinfo
  {author} {\bibfnamefont {R.}~\bibnamefont {Oliveri}}, \bibinfo {author}
  {\bibfnamefont {M.}~\bibnamefont {Orselli}}, \bibinfo {author} {\bibfnamefont
  {A.}~\bibnamefont {Placidi}}, \ and\ \bibinfo {author} {\bibfnamefont
  {J.~E.}\ \bibnamefont {Santos}},\ }\href {\doibase
  10.1088/1475-7516/2022/07/032} {\bibfield  {journal} {\bibinfo  {journal}
  {JCAP}\ }\textbf {\bibinfo {volume} {07}},\ \bibinfo {pages} {032} (\bibinfo
  {year} {2022})},\ \Eprint {http://arxiv.org/abs/2201.11068} {arXiv:2201.11068
  [gr-qc]} \BibitemShut {NoStop}%
\bibitem [{\citenamefont {Camilloni}\ \emph {et~al.}(2023)\citenamefont
  {Camilloni}, \citenamefont {Harmark}, \citenamefont {Orselli},\ and\
  \citenamefont {Rodriguez}}]{Camilloni:2023wyn}%
  \BibitemOpen
  \bibfield  {author} {\bibinfo {author} {\bibfnamefont {F.}~\bibnamefont
  {Camilloni}}, \bibinfo {author} {\bibfnamefont {T.}~\bibnamefont {Harmark}},
  \bibinfo {author} {\bibfnamefont {M.}~\bibnamefont {Orselli}}, \ and\
  \bibinfo {author} {\bibfnamefont {M.~J.}\ \bibnamefont {Rodriguez}},\
  }\href@noop {} {\  (\bibinfo {year} {2023})},\ \Eprint
  {http://arxiv.org/abs/2307.06878} {arXiv:2307.06878 [gr-qc]} \BibitemShut
  {NoStop}%
\bibitem [{\citenamefont {Pei}\ \emph {et~al.}(2016)\citenamefont {Pei},
  \citenamefont {Nampalliwar}, \citenamefont {Bambi},\ and\ \citenamefont
  {Middleton}}]{ref90}%
  \BibitemOpen
  \bibfield  {author} {\bibinfo {author} {\bibfnamefont {G.}~\bibnamefont
  {Pei}}, \bibinfo {author} {\bibfnamefont {S.}~\bibnamefont {Nampalliwar}},
  \bibinfo {author} {\bibfnamefont {C.}~\bibnamefont {Bambi}}, \ and\ \bibinfo
  {author} {\bibfnamefont {M.~J.}\ \bibnamefont {Middleton}},\ }\href {\doibase
  10.1140/epjc/s10052-016-4387-z} {\bibfield  {journal} {\bibinfo  {journal}
  {Eur. Phys. J. C}\ }\textbf {\bibinfo {volume} {76}},\ \bibinfo {pages} {534}
  (\bibinfo {year} {2016})},\ \Eprint {http://arxiv.org/abs/1606.04643}
  {arXiv:1606.04643 [gr-qc]} \BibitemShut {NoStop}%
\bibitem [{\citenamefont {Gou}\ \emph {et~al.}(2010)\citenamefont {Gou},
  \citenamefont {McClintock}, \citenamefont {Steiner}, \citenamefont {Narayan},
  \citenamefont {Cantrell}, \citenamefont {Bailyn},\ and\ \citenamefont
  {Orosz}}]{ref98}%
  \BibitemOpen
  \bibfield  {author} {\bibinfo {author} {\bibfnamefont {L.}~\bibnamefont
  {Gou}}, \bibinfo {author} {\bibfnamefont {J.~E.}\ \bibnamefont {McClintock}},
  \bibinfo {author} {\bibfnamefont {J.~F.}\ \bibnamefont {Steiner}}, \bibinfo
  {author} {\bibfnamefont {R.}~\bibnamefont {Narayan}}, \bibinfo {author}
  {\bibfnamefont {A.~G.}\ \bibnamefont {Cantrell}}, \bibinfo {author}
  {\bibfnamefont {C.~D.}\ \bibnamefont {Bailyn}}, \ and\ \bibinfo {author}
  {\bibfnamefont {J.~A.}\ \bibnamefont {Orosz}},\ }\href {\doibase
  10.1088/2041-8205/718/2/L122} {\bibfield  {journal} {\bibinfo  {journal}
  {Astrophys. J. Lett.}\ }\textbf {\bibinfo {volume} {718}},\ \bibinfo {pages}
  {L122} (\bibinfo {year} {2010})},\ \Eprint {http://arxiv.org/abs/1002.2211}
  {arXiv:1002.2211 [astro-ph.HE]} \BibitemShut {NoStop}%
\bibitem [{\citenamefont {Steiner}\ \emph {et~al.}(2012)\citenamefont
  {Steiner}, \citenamefont {McClintock},\ and\ \citenamefont {Reid}}]{ref100}%
  \BibitemOpen
  \bibfield  {author} {\bibinfo {author} {\bibfnamefont {J.~F.}\ \bibnamefont
  {Steiner}}, \bibinfo {author} {\bibfnamefont {J.~E.}\ \bibnamefont
  {McClintock}}, \ and\ \bibinfo {author} {\bibfnamefont {M.~J.}\ \bibnamefont
  {Reid}},\ }\href {\doibase 10.1088/2041-8205/745/1/L7} {\bibfield  {journal}
  {\bibinfo  {journal} {Astrophys. J. Lett.}\ }\textbf {\bibinfo {volume}
  {745}},\ \bibinfo {pages} {L7} (\bibinfo {year} {2012})},\ \Eprint
  {http://arxiv.org/abs/1111.2388} {arXiv:1111.2388 [astro-ph.HE]} \BibitemShut
  {NoStop}%
\bibitem [{\citenamefont {Steiner}\ \emph {et~al.}(2011)\citenamefont
  {Steiner}, \citenamefont {Reis}, \citenamefont {McClintock}, \citenamefont
  {Narayan}, \citenamefont {Remillard}, \citenamefont {Orosz}, \citenamefont
  {Gou}, \citenamefont {Fabian},\ and\ \citenamefont
  {Torres}}]{steiner2011spin}%
  \BibitemOpen
  \bibfield  {author} {\bibinfo {author} {\bibfnamefont {J.~F.}\ \bibnamefont
  {Steiner}}, \bibinfo {author} {\bibfnamefont {R.~C.}\ \bibnamefont {Reis}},
  \bibinfo {author} {\bibfnamefont {J.~E.}\ \bibnamefont {McClintock}},
  \bibinfo {author} {\bibfnamefont {R.}~\bibnamefont {Narayan}}, \bibinfo
  {author} {\bibfnamefont {R.~A.}\ \bibnamefont {Remillard}}, \bibinfo {author}
  {\bibfnamefont {J.~A.}\ \bibnamefont {Orosz}}, \bibinfo {author}
  {\bibfnamefont {L.}~\bibnamefont {Gou}}, \bibinfo {author} {\bibfnamefont
  {A.~C.}\ \bibnamefont {Fabian}}, \ and\ \bibinfo {author} {\bibfnamefont
  {M.~A.}\ \bibnamefont {Torres}},\ }\href@noop {} {\bibfield  {journal}
  {\bibinfo  {journal} {Monthly Notices of the Royal Astronomical Society}\
  }\textbf {\bibinfo {volume} {416}},\ \bibinfo {pages} {941} (\bibinfo {year}
  {2011})}\BibitemShut {NoStop}%
\bibitem [{\citenamefont {Chen}\ \emph {et~al.}(2016)\citenamefont {Chen},
  \citenamefont {Gou}, \citenamefont {McClintock}, \citenamefont {Steiner},
  \citenamefont {Wu}, \citenamefont {Xu}, \citenamefont {Orosz},\ and\
  \citenamefont {Xiang}}]{ref109}%
  \BibitemOpen
  \bibfield  {author} {\bibinfo {author} {\bibfnamefont {Z.}~\bibnamefont
  {Chen}}, \bibinfo {author} {\bibfnamefont {L.}~\bibnamefont {Gou}}, \bibinfo
  {author} {\bibfnamefont {J.~E.}\ \bibnamefont {McClintock}}, \bibinfo
  {author} {\bibfnamefont {J.~F.}\ \bibnamefont {Steiner}}, \bibinfo {author}
  {\bibfnamefont {J.}~\bibnamefont {Wu}}, \bibinfo {author} {\bibfnamefont
  {W.}~\bibnamefont {Xu}}, \bibinfo {author} {\bibfnamefont {J.}~\bibnamefont
  {Orosz}}, \ and\ \bibinfo {author} {\bibfnamefont {Y.}~\bibnamefont
  {Xiang}},\ }\href {\doibase 10.3847/0004-637X/825/1/45} {\bibfield  {journal}
  {\bibinfo  {journal} {Astrophys. J.}\ }\textbf {\bibinfo {volume} {825}},\
  \bibinfo {pages} {45} (\bibinfo {year} {2016})},\ \Eprint
  {http://arxiv.org/abs/1601.00615} {arXiv:1601.00615 [astro-ph.HE]}
  \BibitemShut {NoStop}%
\bibitem [{\citenamefont {{Shafee}}\ \emph {et~al.}(2006)\citenamefont
  {{Shafee}}, \citenamefont {{McClintock}}, \citenamefont {{Narayan}},
  \citenamefont {{Davis}}, \citenamefont {{Li}},\ and\ \citenamefont
  {{Remillard}}}]{Shafee06}%
  \BibitemOpen
  \bibfield  {author} {\bibinfo {author} {\bibfnamefont {R.}~\bibnamefont
  {{Shafee}}}, \bibinfo {author} {\bibfnamefont {J.~E.}\ \bibnamefont
  {{McClintock}}}, \bibinfo {author} {\bibfnamefont {R.}~\bibnamefont
  {{Narayan}}}, \bibinfo {author} {\bibfnamefont {S.~W.}\ \bibnamefont
  {{Davis}}}, \bibinfo {author} {\bibfnamefont {L.-X.}\ \bibnamefont {{Li}}}, \
  and\ \bibinfo {author} {\bibfnamefont {R.~A.}\ \bibnamefont {{Remillard}}},\
  }\href {\doibase 10.1086/498938} {\bibfield  {journal} {\bibinfo  {journal}
  {Astrophys. J. Lett}\ }\textbf {\bibinfo {volume} {636}},\ \bibinfo {pages}
  {L113} (\bibinfo {year} {2006})},\ \Eprint
  {http://arxiv.org/abs/astro-ph/0508302} {astro-ph/0508302} \BibitemShut
  {NoStop}%
\bibitem [{\citenamefont {McClintock}\ \emph {et~al.}(2006)\citenamefont
  {McClintock}, \citenamefont {Shafee}, \citenamefont {Narayan}, \citenamefont
  {Remillard}, \citenamefont {Davis},\ and\ \citenamefont {Li}}]{ref118}%
  \BibitemOpen
  \bibfield  {author} {\bibinfo {author} {\bibfnamefont {J.~E.}\ \bibnamefont
  {McClintock}}, \bibinfo {author} {\bibfnamefont {R.}~\bibnamefont {Shafee}},
  \bibinfo {author} {\bibfnamefont {R.}~\bibnamefont {Narayan}}, \bibinfo
  {author} {\bibfnamefont {R.~A.}\ \bibnamefont {Remillard}}, \bibinfo {author}
  {\bibfnamefont {S.~W.}\ \bibnamefont {Davis}}, \ and\ \bibinfo {author}
  {\bibfnamefont {L.-X.}\ \bibnamefont {Li}},\ }\href {\doibase 10.1086/508457}
  {\bibfield  {journal} {\bibinfo  {journal} {Astrophys. J.}\ }\textbf
  {\bibinfo {volume} {652}},\ \bibinfo {pages} {518} (\bibinfo {year}
  {2006})},\ \Eprint {http://arxiv.org/abs/astro-ph/0606076}
  {arXiv:astro-ph/0606076} \BibitemShut {NoStop}%
\bibitem [{\citenamefont {{Middleton}}\ \emph {et~al.}(2014)\citenamefont
  {{Middleton}}, \citenamefont {{Miller-Jones}},\ and\ \citenamefont
  {{Fender}}}]{ref94}%
  \BibitemOpen
  \bibfield  {author} {\bibinfo {author} {\bibfnamefont {M.~J.}\ \bibnamefont
  {{Middleton}}}, \bibinfo {author} {\bibfnamefont {J.~C.~A.}\ \bibnamefont
  {{Miller-Jones}}}, \ and\ \bibinfo {author} {\bibfnamefont {R.~P.}\
  \bibnamefont {{Fender}}},\ }\href {\doibase 10.1093/mnras/stu056} {\bibfield
  {journal} {\bibinfo  {journal} {mnras}\ }\textbf {\bibinfo {volume} {439}},\
  \bibinfo {pages} {1740} (\bibinfo {year} {2014})},\ \Eprint
  {http://arxiv.org/abs/1401.1829} {arXiv:1401.1829 [astro-ph.HE]} \BibitemShut
  {NoStop}%
\bibitem [{\citenamefont {Tchekhovskoy}\ \emph {et~al.}(2011)\citenamefont
  {Tchekhovskoy}, \citenamefont {Narayan},\ and\ \citenamefont
  {McKinney}}]{Tchekhovskoyetal.2011}%
  \BibitemOpen
  \bibfield  {author} {\bibinfo {author} {\bibfnamefont {A.}~\bibnamefont
  {Tchekhovskoy}}, \bibinfo {author} {\bibfnamefont {R.}~\bibnamefont
  {Narayan}}, \ and\ \bibinfo {author} {\bibfnamefont {J.~C.}\ \bibnamefont
  {McKinney}},\ }\href {\doibase 10.1111/j.1745-3933.2011.01147.x} {\bibfield
  {journal} {\bibinfo  {journal} {Monthly Notices of the Royal Astronomical
  Society: Letters}\ }\textbf {\bibinfo {volume} {418}},\ \bibinfo {pages}
  {L79} (\bibinfo {year} {2011})}\BibitemShut {NoStop}%
\bibitem [{\citenamefont {Russell}\ \emph {et~al.}(2013)\citenamefont
  {Russell}, \citenamefont {Gallo},\ and\ \citenamefont
  {Fender}}]{Russell2013ws}%
  \BibitemOpen
  \bibfield  {author} {\bibinfo {author} {\bibfnamefont {D.~M.}\ \bibnamefont
  {Russell}}, \bibinfo {author} {\bibfnamefont {E.}~\bibnamefont {Gallo}}, \
  and\ \bibinfo {author} {\bibfnamefont {R.~P.}\ \bibnamefont {Fender}},\
  }\href {\doibase 10.1093/mnras/stt176} {\bibfield  {journal} {\bibinfo
  {journal} {Mon. Not. Roy. Astron. Soc.}\ }\textbf {\bibinfo {volume} {431}},\
  \bibinfo {pages} {405} (\bibinfo {year} {2013})},\ \Eprint
  {http://arxiv.org/abs/1301.6771} {arXiv:1301.6771 [astro-ph.HE]} \BibitemShut
  {NoStop}%
\end{thebibliography}%

\end{document}